%% file: main.tex
\documentclass{iopjournal}
\usepackage{lmodern}
\usepackage{longtable}
\usepackage{amsmath}
\usepackage{amssymb}
\usepackage{multirow}
\usepackage[table]{xcolor}
\usepackage{graphicx}
\usepackage{url}

\renewcommand{\headrulewidth}{0pt}
\renewcommand{\articletype}[1]{}
\renewcommand{\affil}[1]{\par{\fontsize{8}{10}\selectfont\raggedright #1\par}}

\usepackage[section]{placeins}


\begin{document}

\articletype{Paper}

\title{Direct polarisation-state reconstruction with a temporal pointer: accessible states, resolution and optimal coupling}

\author{S. P. Gervais$^1$ and Lambert Giner$^{2,*}$}
\affil{$^1$Department of Astronomy \& Physics, Saint Mary's University, Halifax, B3H 3C3, NS, Canada}
\affil{$^2$D\'epartement de physique et astronomie, Universit\'e de Moncton, Moncton, New Brunswick, Canada}
\affil{$^{*}$ Author to whom any correspondence should be addressed.}
\email{lambert.giner@umoncton.ca}

\begin{abstract}
Weak measurements are used to amplify small signals, to address foundational
questions, and to characterise quantum states without tomography. In direct
state reconstruction each amplitude of the state vector is proportional to a
single complex weak value, which for a temporal pointer is read from the mean
arrival time and the mean frequency of the postselected pulse. Such an
experiment requires three quantities to be fixed, the coupling, the angular
resolution demanded of the reconstruction, and the set of states over which
the method remains usable, and these are commonly set by the available
apparatus rather than derived. We derive all three for a Gaussian temporal
pointer and an arbitrary postselection state. The states that cannot be
reconstructed form a spherical cap centred on the state orthogonal to the
postselection, whose geodesic half-angle follows from the largest of three
independent thresholds and involves no adjustable parameter. Two states are
distinguishable only if they differ by more than a minimal step fixed by the
coupling, by the number of incident photons and by the sensitivity of the weak
value along the path followed. For an inversion at first order in the coupling
the systematic and statistical errors are opposed and fix an optimal coupling
that varies as the inverse sixth root of the number of incident photons,
whereas an exact inversion at finite coupling removes the systematic error, the remaining bound being one on the two-centroid readout rather than on the inversion, since below an amplitude equal to a quarter of the coupling two states give the same pair of centroids while their profiles differ. These results allow a direct reconstruction experiment to be
dimensioned before the apparatus is assembled.
\end{abstract}



\input{section1_introduction}


\input{section2_temporal_pointer}


\input{section3_statistical_analysis}


\input{section4_angular_resolution}


\input{section5_optimal_coupling}


\input{section6_conclusion}

\bibliographystyle{iopart-num}
\bibliography{references}

\section*{Acknowledgments}
L.G. acknowledges support from the Natural Sciences and Engineering Research
Council of Canada (RGPIN-2022-04394).

During the preparation of this manuscript, the authors used Claude
(Anthropic, Opus 5) to edit human-written text, including improvements to
language, clarity and structure, to generate figures from data produced by the
authors, and to support the literature review. All references were checked by
the authors against the original sources. The authors critically revised all
output and take full responsibility for the accuracy, integrity and
originality of the content.

\section*{Data availability statement}
All data that support the findings of this study are included within the article (and any supplementary files).

\section*{Conflict of interest}
The authors declare no conflict of interest.

\section*{ORCID iDs}
S. P. Gervais \orcid{0009-0003-1815-8156} \\
\noindent Lambert Giner \orcid{0000-0002-4412-1248}

\end{document}

%% file: section1_introduction.tex

\section{Introduction}
\label{sec:intro}

Determining the state of a quantum system is central to the development of
quantum technologies. The standard method, quantum state
tomography~\cite{Altepeter2004}, performs a global inversion of a complete
data set~\cite{James2001,DAriano2003}, so that no individual outcome
corresponds to a coefficient of the state.

Weak measurement provides an alternative, in which the system under
investigation is not measured directly but weakly coupled to a pointer, so
that a modification of the pointer carries the information on the state.
Weak measurements have been used to amplify small
signals~\cite{Hosten2008,Dixon2009}, to address foundational
questions~\cite{Pusey2014}, and to characterise quantum
states~\cite{LundeenBamber2012}; these applications are reviewed
in~\cite{Dressel2014}. The weak value of an observable $A$, for a system
prepared in an initial state $|\psi\rangle$ and postselected on a state
$|\varsigma\rangle$, is
\begin{equation}
\langle A\rangle_W
=\frac{\langle\varsigma|A|\psi\rangle}{\langle\varsigma|\psi\rangle},
\label{eq:weakvalue}
\end{equation}
which is in general complex~\cite{AAV1988}. It is moreover unbounded, since
the denominator vanishes when the postselection state is orthogonal to the
initial state. The coupling between the system under investigation and the
pointer is weak enough for a weak measurement to leave the state unaffected
to first order. For a pointer with a Gaussian envelope, the real and
imaginary parts of the weak value are encoded respectively in the mean
position and the mean momentum of the pointer~\cite{Jozsa2007}.

Direct state reconstruction from weak values~\cite{Lundeen2011} relies on
each amplitude coefficient of the state vector being proportional to a
single complex weak value, up to a multiplicative constant that is removed
by normalisation. The state is therefore obtained directly from the reading
of an experimental quantity, without global inversion. The scheme has been
used to reconstruct the state vector of a polarisation qubit with a spatial
pointer~\cite{Salvail2013}.

In a temporal implementation the pointer is the envelope of a light pulse of
width $\sigma$, so that the mean arrival time is the pointer position and the
frequency its conjugate variable. A delay $\tau$ applied to the horizontal
polarisation component $|H\rangle$ then encodes the real part of the weak
value $\mathrm{Re}(\langle A\rangle_W)$ in the temporal centroid of the
postselected pulse, that is its mean arrival time, and the imaginary part
$\mathrm{Im}(\langle A\rangle_W)$ in the spectral centroid. This
configuration has been used to amplify small phase
shifts~\cite{BrunnerSimon2010} and to exploit technical
noise~\cite{Kedem2012}, where a single coupling $x=\tau/\sigma$ is a fixed
property of the apparatus rather than a quantity to be optimised. In a
spatial implementation the pointer is the transverse spatial mode itself,
which is altered when it is displaced in free space by walk-off
crystals~\cite{Thekkadath2016}. The temporal implementation leaves the
spatial mode unaltered, and is therefore compatible with integration in a
single-mode waveguide.

The two centroids, temporal and spectral, are carried by the same
postselected state and are determined in two detection configurations, which
share the same optical path and differ only in the detector. A temporal
pointer has a single coordinate, whose conjugate variable is its Fourier
transform, so that the two centroids cannot be recorded simultaneously. The
transverse mode, by contrast, has two independent coordinates, and both
components have been read simultaneously in a single
configuration~\cite{Kobayashi2014,Hariri2019}.

The experimenter must nevertheless fix three quantities: the set of states on
which the reconstruction method can be used, which is limited because the
weak value diverges as the postselection state approaches orthogonality with
the initial state; the angular resolution required of the reconstruction
method; and the coupling $x=\tau/\sigma$. In the references cited above these
three quantities are fixed by the available experimental arrangement, whereas
here we establish them.

The weak regime is imposed here by the direct reconstruction itself, and is not adopted for a metrological advantage. Postselection provides no advantage in total resources~\cite{FerrieCombes2014,KneeGauger2014,Jordan2014}. What we derive is the precision attainable under that constraint. The value of the coupling can be chosen freely in an experiment, but not arbitrarily. It must be small enough for the first-order inversion to remain accurate, whereas the shifts it produces on the pointer must be large enough to exceed the uncertainty on the centroids. The two requirements are opposed and fix an optimal value.

In section~\ref{sec:pointer} we develop the framework for direct
reconstruction of the state vector with a temporal pointer, which gives the
two centroids at arbitrary coupling and reconstruction formulae valid for an
arbitrary postselection. On that basis we answer three questions. The first
is which states the method can reconstruct, since the divergence of the weak
value excludes those close to the state orthogonal to the postselection
state, and it is treated in section~\ref{sec:errors}. The second is how close
two states can be while the reconstruction still distinguishes them, and it
is treated in section~\ref{sec:resolution}. The third is how the state can
still be recovered when the coupling is not small, and where reading the two
centroids alone ceases to be enough, both being treated in
section~\ref{sec:optimal}.

%% file: section2_temporal_pointer.tex

\section{Weak measurement with a temporal pointer}
\label{sec:pointer}

The weak limit is the regime in which the state-reconstruction formulae
take their simplest form, but it is not necessarily the regime in which
the measurement is performed. We therefore derive exact expressions for
the temporal and spectral centroids of a postselected light pulse before
imposing the weak-limit constraint. These two centroids give the two
quadratures of the weak value simultaneously. The state is then
reconstructed by an algebraic inversion, and the expressions obtained at
non-zero coupling provide the basis for sections~\ref{sec:errors}
and~\ref{sec:optimal}.

\subsection{Setup and interaction}
\label{sec:setup}

\begin{figure}[t]
\centering
\includegraphics[width=\linewidth]{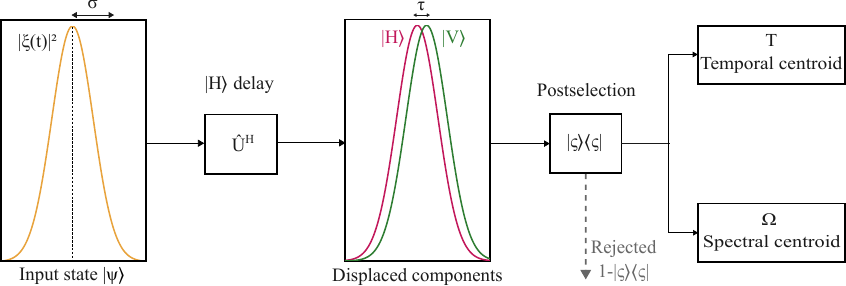}
\caption{Conceptual scheme of the temporal weak measurement. A pulse of
temporal width $\sigma$ enters in an unknown polarisation state
$|\psi\rangle$. The operator $\hat{U}^{H}$ delays the $|H\rangle$
component by $\tau$, leaving $|V\rangle$ unchanged. Projection onto
$|\varsigma\rangle$ recombines the two components into a single pulse,
the remaining photons being discarded. The temporal centroid $T$ and the
spectral centroid $\Omega$ of the postselected pulse are recorded in two
separate detection configurations. The scheme is not to scale.}
\label{fig:scheme}
\end{figure}

Consider a system consisting of a quantum state to be characterised,
coupled to a second quantum state referred to as the pointer. In this
work we consider polarisation states prepared in the general form
$|\psi\rangle=a|H\rangle+b|V\rangle$, where $a$ is real by a
choice of global phase and $b$ is complex, with $|a|^2+|b|^2=1$. This
qubit is coupled to a temporal pointer
$\xi(t)=\left(2\pi\sigma^2\right)^{-1/4}e^{-t^2/4\sigma^2}$, the
normalised Gaussian envelope of a laser pulse, where $\sigma$ denotes the
standard deviation of the intensity distribution $|\xi(t)|^2$. The
measurement chain is sketched in figure~\ref{fig:scheme}.

The coupling between the system and the pointer is realised by the von
Neumann unitary operator
$\hat U^{H}=\exp\!\left(-i\tau\,|H\rangle\langle H|\otimes\hat p_t\right)$
\cite{vonNeumann1932}, where $\hat p_t=-i\,\mathrm{d}/\mathrm{d}t$ is the
generator of temporal translations: the pointer attached to the
$|H\rangle$ component is displaced by an amount
$\tau$ under the action of the operator, while the pointer attached to
the $|V\rangle$ component is left unchanged. Since the exponential is not
expanded, all the expressions that follow remain valid at arbitrary
coupling. The coupling strength is measured by the dimensionless ratio
$x=\tau/\sigma$.

The weak value is not read directly on the polarisation state to be
measured, but on the change in the pointer state after the von Neumann
operator has acted and after the polarisation state has been projected
onto a final state $|\varsigma\rangle=\mu|H\rangle+\nu|V\rangle$, where
$\mu$ and $\nu$ are complex numbers with $|\mu|^2+|\nu|^2=1$. This
projection leaves the pointer in the superposition
\begin{equation}
F(t)=A\,\xi(t-\tau)+B\,\xi(t)
\label{eq:Ft}
\end{equation}
of two identical Gaussian components, one displaced by $\tau$ and the
other not. Their weights are set by the states $|\psi\rangle$ and
$|\varsigma\rangle$: the coefficient of the displaced component is
$A=a\bar\mu$ and that of the undisplaced component is $B=b\bar\nu$. The
overlap integral between the two components is given by
\begin{equation}
q=\int_{-\infty}^{\infty}\xi(t-\tau)\,\xi(t)\,dt=e^{-x^2/8},
\label{eq:q}
\end{equation}
which equals unity at $x=0$ and decreases as $x$ increases. The
postselection has thus transferred the polarisation amplitudes onto the
weights of the two pointer components. These individual weights define
the weak value of the projector on $|H\rangle$,
\begin{equation}
W=\frac{A}{A+B}
=\frac{\langle\varsigma|H\rangle\langle H|\psi\rangle}
{\langle\varsigma|\psi\rangle},
\label{eq:W}
\end{equation}
which is equation~\eqref{eq:weakvalue}~\cite{AAV1988} written for the
present protocol.
The centroids of $F(t)$ obtained in section~\ref{sec:shifts} therefore
give access to $W$.

The norm of the postselected pointer state defined in
equation~\eqref{eq:Ft} is the probability that a photon survives
postselection, and is given by
\begin{equation}
\mathcal{N}^2=|A|^2+|B|^2+2\,\mathrm{Re}(A\bar B)\,q .
\label{eq:N2}
\end{equation}
In this expression the overlap integral $q$ modulates the interference
term alone. Of the $N_{\rm total}$ photons incident on the setup, only
$N=\mathcal{N}^2N_{\rm total}$ survive postselection and reach the
detector, the remainder being discarded as shown in
figure~\ref{fig:scheme}. A small $\mathcal{N}^2$ leaves few photons at the detector and
therefore sets the statistical precision of the measurement, analysed in
section~\ref{sec:errors}. In the weak-measurement regime $x\to0$,
$\mathcal{N}^2$ equals $|\langle\varsigma|\psi\rangle|^2$. For $x>0$, $q$
is smaller than unity, which imposes a lower bound on $\mathcal{N}^2$, as
shown in section~\ref{sec:floor}. The same departure of $q$ from unity
provides the expansion parameter of section~\ref{sec:optimal}.

\subsection{Temporal and spectral shifts}
\label{sec:shifts}

The temporal centroid, or mean arrival time $T$, and the spectral
centroid, or mean frequency shift $\Omega$, are the first moments of the
postselected pointer state in the two conjugate domains,
\begin{equation}
T=\frac{1}{\mathcal{N}^2}\int_{-\infty}^{\infty}t\,|F(t)|^2\,dt ,
\qquad
\Omega=\frac{1}{\mathcal{N}^2}\int_{-\infty}^{\infty}
\omega\,|\tilde F(\omega)|^2\,d\omega ,
\label{eq:centroids}
\end{equation}
where $\tilde F$ is the Fourier transform of $F$. They are obtained
independently, starting with the temporal one. Evaluating the first
integral gives
\begin{equation}
T=\tau\,\frac{|A|^2+\mathrm{Re}(A\bar B)\,q}{\mathcal{N}^2},
\label{eq:Texact}
\end{equation}
which equals $\tau\,\mathrm{Re}(W)$ in the weak-measurement regime. The
second integral is evaluated in the conjugate domain, where the pointer
is
\begin{equation}
\tilde\xi(\omega)=\left(\frac{2\sigma^2}{\pi}\right)^{1/4}
e^{-\sigma^2\omega^2},
\label{eq:xitilde}
\end{equation}
normalised in frequency. The shift theorem
turns the displaced component into a phase factor, and the power spectrum
acquires cosine and sine cross terms:
\begin{equation}
|\tilde F(\omega)|^2
=\left(\frac{2\sigma^2}{\pi}\right)^{1/2}
\Big[|A|^2+|B|^2+2\,\mathrm{Re}(A\bar B)\cos\omega\tau
+2\,\mathrm{Im}(A\bar B)\sin\omega\tau\Big]e^{-2\sigma^2\omega^2}.
\label{eq:spectrum}
\end{equation}
Only the sine term contributes to $\Omega$. The corresponding integral is
obtained by differentiating with respect to the delay the auxiliary
Gaussian integral
\begin{equation}
G(\tau)=\sqrt{\frac{\pi}{2\sigma^2}}\;e^{-\tau^2/8\sigma^2}
=\sqrt{\frac{\pi}{2\sigma^2}}\;q .
\label{eq:Gtau}
\end{equation}
The exact result is
\begin{equation}
\Omega=\frac{\tau}{2\sigma^2}\,
\frac{\mathrm{Im}(A\bar B)\,q}{\mathcal{N}^2},
\label{eq:Omexact}
\end{equation}
which equals $(\tau/2\sigma^2)\,\mathrm{Im}(W)$ in the weak-measurement
regime.

As an illustration, take as the state to be characterised the
right-circular polarisation state
$|\psi\rangle=|R\rangle=(|H\rangle+i|V\rangle)/\sqrt2$, postselected on
the diagonal polarisation state
$|\varsigma\rangle=|D\rangle=(|H\rangle+|V\rangle)/\sqrt2$. Evaluating
equations~\eqref{eq:Texact} and~\eqref{eq:Omexact} at $q=1$ gives
$T=\tau/2$ and $\Omega=-\tau/(4\sigma^2)$. Any state with
$\mathrm{Im}(b)>0$ thus gives a negative mean frequency shift, the sign
being fixed by the Fourier convention adopted here. Inverting the
relations obtained above in the weak-measurement regime yields
$\mathrm{Re}(W)=T/\tau=1/2$ and
$\mathrm{Im}(W)=(2\sigma^2/\tau)\,\Omega=-1/2$. The two centroids,
temporal and spectral, obtained from the same postselected light pulse
therefore determine $\mathrm{Re}(W)$ and $\mathrm{Im}(W)$
completely~\cite{Jozsa2007}, provided the coefficients $\mu$ and $\nu$
are both non-zero.

\subsection{Direct state reconstruction}
\label{sec:reconstruction}

Once both centroids are known, the state can be reconstructed. They enter
the definition of a quantity
\begin{equation}
X=\frac{T}{\tau}+i\,\frac{2\sigma^2}{\tau}\,\Omega ,
\label{eq:X}
\end{equation}
which may be complex and which tends to $W$ in the weak-value regime. The
state then follows from an algebraic inversion of this quantity through
the relations
\begin{equation}
\frac{b}{a}=\frac{\bar\mu\,(1-X)}{\bar\nu\,X},
\qquad
a=\frac{1}{\sqrt{1+|b/a|^2}},
\qquad
b=a\,\frac{b}{a} .
\label{eq:reconstruction}
\end{equation}
The inversion holds for an arbitrary state to be determined and for any
postselection state, provided $\mu$ and $\nu$ are both non-zero, and is
exact at $x=0$. It determines the state up to a global phase: the third relation returns
the representative with $a$ positive, whereas a path on the Poincar\'e
sphere may reach the same state with $a$ negative. Both representatives
give the same $\mathcal{N}^2$ and the same $X$, and the reconstruction
error is evaluated in the representative used to parametrise the path.

Applying this formula with postselection on the diagonal state
$|\varsigma\rangle=|D\rangle$ for several input states:
$|\psi\rangle=|H\rangle$ gives $X=1$, hence $a=1$ and $b=0$;
$|\psi\rangle=|R\rangle=(|H\rangle+i|V\rangle)/\sqrt2$ gives
$X=(1-i)/2$, hence $a=1/\sqrt2$ and $b=i/\sqrt2$; and
$|\psi\rangle=|V\rangle$ gives $X=0$, hence $a=0$ and $b=1$. A
singularity does appear, however, when the input state is orthogonal to
the postselection state, that is when
$|\psi\rangle=|\varsigma_\perp\rangle$. In that case the postselection
probability vanishes, no photon reaches the detector, and $X$ is
undefined. This case is analysed in section~\ref{sec:errors}. So far $X$
has been defined through equation~\eqref{eq:X}, which assumes the
weak-value regime. This expression can be generalised to arbitrary
coupling $x$: for $|\varsigma\rangle=|D\rangle$ it takes the form
\begin{equation}
X(q)=\frac{a^2+(R+iJ)\,q}{1+2Rq},
\label{eq:Xq}
\end{equation}
with $R=\mathrm{Re}(a\bar b)$ and $J=\mathrm{Im}(a\bar b)$, which is the
starting point of section~\ref{sec:optimal}.

The reconstruction given by equation~\eqref{eq:reconstruction} is exact
at zero coupling ($x=0$) and holds for any input state. It contains no
information, however, on the precision attainable at non-zero coupling.
Section~\ref{sec:errors} addresses this question.

%% file: section3_statistical_analysis.tex

\section{Validity and precision of the reconstruction}
\label{sec:errors}

The state reconstruction of section~\ref{sec:reconstruction} fails for an
input state $|\psi\rangle$ orthogonal to the postselection state
$|\varsigma\rangle$, since no photon reaches the detector and $X$ is
therefore undefined. It is also degraded for the states neighbouring that
orthogonal state, where the photon flux is low and $X$ departs from $W$.
This section determines the extent of the region in which the
reconstruction cannot be used, and the precision reached outside that
exclusion region, for the postselection $|\varsigma\rangle=|D\rangle$ and
for the couplings $x=0.01$ and $x=0.10$.

\subsection{The exclusion region}
\label{sec:cap}

The weak value of equation~\eqref{eq:W} is a ratio whose denominator is
the postselection amplitude itself. That amplitude,
$\langle\varsigma|\psi\rangle=A+B$, where $A=\bar\mu a$ and $B=\bar\nu b$
are the amplitudes of the temporal pointer superposition defined in
equation~\eqref{eq:Ft}, vanishes when
$|\psi\rangle=|\varsigma_\perp\rangle$, and the weak value therefore
diverges at that state. For postselection on the diagonal state
$|\varsigma\rangle=|D\rangle$, the orthogonal state is the antidiagonal
state $|\varsigma_\perp\rangle=|A\rangle$. Writing
$\varepsilon=|\langle\varsigma|\psi\rangle|$ for the modulus of the
postselection amplitude, $|W|$ grows as $1/\varepsilon$ as $\varepsilon$
tends to zero, whereas the postselection probability $\mathcal{N}^2$ of
equation~\eqref{eq:N2} equals $\varepsilon^2$ in the ideal weak regime
($q=1$) and therefore falls quadratically. The modulus of the weak value
and the photon flux thus follow different powers of $\varepsilon$ and
opposite trends. This mismatch of exponents is not the origin of the
failure of the reconstruction. That origin is the existence of a lower
limit on $\varepsilon$, which depends on the photon number, on the
detector and on the validity of the weak-regime approximation, as
established in section~\ref{sec:floor}.

For an arbitrary state $|\psi\rangle=a|H\rangle+b|V\rangle$, the
postselection probability at $q=1$ is
\begin{equation}
\mathcal{N}^2=\frac{1+2a\,\mathrm{Re}(b)}{2}
\label{eq:N2gen}
\end{equation}
and depends only on the real part of the coefficient $b$, since neither
$\mathrm{Im}(b)$ nor $|b|$ appears. A pure polarisation state can also be
written as a Stokes vector with coordinates $S_1=a^2-|b|^2$,
$S_2=2a\,\mathrm{Re}(b)$ and $S_3=2a\,\mathrm{Im}(b)$, so that the
postselection probability becomes
\begin{equation}
\mathcal{N}^2=\frac{1+S_2}{2},
\label{eq:N2S2}
\end{equation}
which depends on a single one of the three Stokes parameters. Since
$|D\rangle$ and $|A\rangle$ are orthogonal, they are antipodal on the
Poincar\'e sphere, with coordinates $S_2=1$ and $S_2=-1$ respectively, so
the geodesic angle $\alpha$ between $|\psi\rangle$ and $|A\rangle$ obeys
$\cos\alpha=-S_2$ and equation~\eqref{eq:N2S2} becomes
$\mathcal{N}^2=\sin^2(\alpha/2)$. The reconstruction requires the
postselection amplitude to exceed a minimum value $\varepsilon_{\rm min}$,
which is determined in section~\ref{sec:floor}. The condition
$\mathcal{N}\ge\varepsilon_{\rm min}$ therefore excludes the states
$|\psi\rangle$ whose angle $\alpha$ from $|A\rangle$ is smaller than
$\alpha_{\rm min}$, defined by
\begin{equation}
\cos\alpha_{\rm min}=1-2\varepsilon_{\rm min}^2 .
\label{eq:cap}
\end{equation}

\begin{figure}[htbp]
\centering
\includegraphics[width=0.75\linewidth]{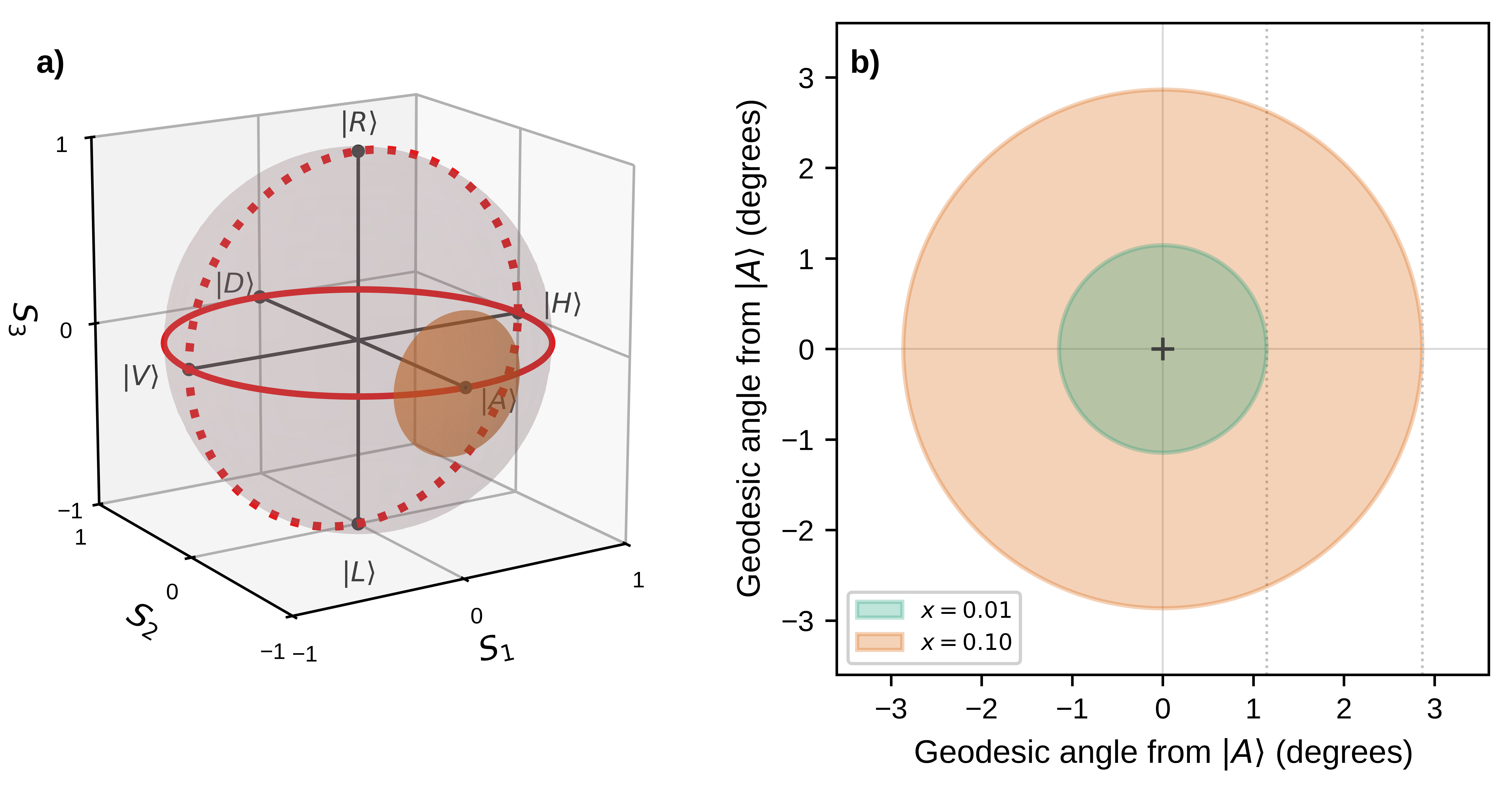}
\caption{\label{fig:cap}Excluded region on the Poincar\'e sphere for
postselection on $|D\rangle$. (a) The cap of geodesic half-angle
$\alpha_{\rm min}$ centred on the antidiagonal state $|A\rangle$, drawn for
$x=0.10$, with the six poles of the sphere marked. The two paths of
section~\ref{sec:examples} are drawn in red: the great circle $S_3=0$ of
linear polarisation states as a solid line, which crosses the cap at
$|A\rangle$, and the great circle $S_2=0$ as a dotted line. (b) The two caps
seen from $|A\rangle$, of half-angle $1.1^\circ$ at $x=0.01$ and $2.9^\circ$ at
$x=0.10$, the factor between the two radii following from the change of
limiting floor in equation~\eqref{eq:epsmin}. Parameters
$N_{\rm min}=10$, $\Gamma=100$~s$^{-1}$, $t_{\rm acq}=5$~s,
$N_{\rm total}=5\times10^{6}$.}
\end{figure}

Since the postselection probability on $|D\rangle$ depends only on the
Stokes component $S_2$, the region of excluded states $|\psi\rangle$ is a
spherical cap centred on $|A\rangle$, shown in figure~\ref{fig:cap} for
the two couplings: a linear state and an elliptical state characterised by
the same geodesic angle $\alpha$ are equally accessible. For the
particular case of linear states ($\mathrm{Im}(b)=0$),
$|\psi\rangle=\cos\theta\,|H\rangle+\sin\theta\,|V\rangle$ with $\theta$
the physical angle, the postselection probability can be written
\begin{equation}
\mathcal{N}^2=\frac{1+\sin 2\theta}{2},
\label{eq:N2lin}
\end{equation}
which serves as the foundation of section~\ref{sec:resolution}. The shape
of the forbidden region has a geometrical interpretation: up to a scale
factor, the weak value is the stereographic projection of the Poincar\'e
sphere from $|A\rangle$ onto the plane $S_2=0$ \cite{Kobayashi2014}. The
excluded states are therefore the neighbours of the pole of that
projection, as shown in section~\ref{sec:proj}.

\subsection{Propagation of the centroid uncertainties}
\label{sec:stat}

Section~\ref{sec:cap} determined where the state reconstruction is
possible, and the present section establishes its precision, which depends
on the two centroids, the mean arrival time $T$ and the mean frequency
shift $\Omega$ defined in section~\ref{sec:shifts}. Each centroid is a
sample mean, so its error is the width of its distribution divided by the
square root of the number of postselected events. In the weak-value regime
those widths are the widths of $|\xi(t)|^2$ and of
$|\tilde\xi(\omega)|^2$, that is $\sigma$ and $1/(2\sigma)$, and the
uncertainties are therefore
\begin{subequations}
\begin{eqnarray}
\delta T=\frac{\sigma}{\sqrt N},
\label{eq:dT}\\
\delta\Omega=\frac{1}{2\sigma\sqrt N},
\label{eq:dOm}
\end{eqnarray}
\end{subequations}
where
\begin{equation}
N=\mathcal{N}^2N_{\rm total}
\label{eq:Nps}
\end{equation}
is the number of postselected photons. Since $T$ and $\Omega$ are measured
in different experimental configurations, they are statistically
independent: the time-bandwidth product sets a minimum value on the
product of the two distribution widths, but imposes no constraint on the
correlation between two means sampled separately. These uncertainties fix
the precision of the reconstruction and determine, in
section~\ref{sec:floor}, whether statistical noise limits the approach to
$|A\rangle$.

The two measured centroids determine the quantity $X$ defined in
equation~\eqref{eq:X}, which equals the weak value $W$ in the weak regime,
and whose real and imaginary parts are written $W_{\rm Re}$ and $W_{\rm
Im}$. For postselection on the diagonal state
$|\varsigma\rangle=|D\rangle$, the projectors $\pi_H$ and $\pi_V$ have
weak values $W_H$ and $W_V$, with $W_H$ the quantity $W$ defined in
equation~\eqref{eq:W}. Since the two projectors sum to the identity,
$W_H+W_V=1$, and a single measurement of $X$ therefore determines both
weak values. The reconstruction equations introduced in
section~\ref{sec:reconstruction} can be rewritten as
\begin{equation}
\psi_H=vW_H,\qquad \psi_V=v(1-W_H),
\label{eq:direct}
\end{equation}
where $v$ is a common constant, so both amplitudes are determined by a
single measured quantity, up to that constant, which normalisation
removes. The error propagation is then immediate: with $x=\tau/\sigma$,
the uncertainties $\delta T$ and $\delta\Omega$ give
\begin{equation}
|\delta X|^2=\left(\frac{\delta T}{\tau}\right)^2
+\left(\frac{2\sigma^2\delta\Omega}{\tau}\right)^2
=\frac{2\sigma^2}{\tau^2N}=\frac{2}{x^2N},
\label{eq:dX}
\end{equation}
and that uncertainty in turn gives the uncertainty on the non-normalised
amplitudes. The temporal and spectral uncertainties contribute with equal
weight, since the pointer is Gaussian and saturates the time-bandwidth
product. Substituting equation~\eqref{eq:Nps} into equation~\eqref{eq:dX}
shows that $\mathcal{N}$ enters the uncertainty only through the number of
postselected photons,
\begin{equation}
|\delta X|=\frac{\sqrt2}{x\,\mathcal{N}\sqrt{N_{\rm total}}} .
\label{eq:dXeps}
\end{equation}

The relevant quantity is the relative uncertainty on the weak value. In
the weak regime, the definitions of $W_H$ and $\mathcal{N}$ give the
modulus
\begin{equation}
|W_H|=\frac{a}{\sqrt2\,\mathcal{N}},
\label{eq:Wmod}
\end{equation}
and with equation~\eqref{eq:dXeps} $\mathcal{N}$ cancels, so that the
ratio reads
\begin{equation}
\frac{|\delta X|}{|W_H|}=\frac{2}{a\,x\sqrt{N_{\rm total}}} .
\label{eq:rel}
\end{equation}
This ratio alone determines the precision of the reconstruction, without
any separate propagation for the real and imaginary parts of the weak
value and without any assumption on the state $|\psi\rangle$, and it
depends only on $a$, on $x$ and on $N_{\rm total}$, and not on the
postselection amplitude. Since the non-normalised amplitudes $\psi_H$ and
$\psi_V$ are given by equation~\eqref{eq:direct}, the relative uncertainty
on $\psi_H$ is that same ratio. Normalisation involves both amplitudes, so
the uncertainty on $a$ also depends on $b$, and the propagation onto the
normalised amplitudes is carried out in section~\ref{sec:optimal}. The
inverse of the ratio~\eqref{eq:rel} is the signal-to-noise ratio of the
measured quantity $X$,
\begin{equation}
\mathrm{SNR}_X=\frac{a\,x\sqrt{N_{\rm total}}}{2},
\label{eq:SNRX}
\end{equation}
in which the dependence on $b$ has disappeared, since the amplification of
$|W_H|$ and the reduction of $N$ both depend on $|a+b|$. The minimum
number of incident photons therefore depends only on the numerator of the
weak value, and not on the postselection amplitude \cite{Kofman2012}. If
only the mean arrival time $T$ is measured, the signal-to-noise ratio
becomes $\mathrm{SNR}_T$, and the ratio of the two follows,
\begin{subequations}
\begin{eqnarray}
\mathrm{SNR}_T=x\,|W_{\rm Re}|\sqrt N,
\label{eq:SNRT}\\
\frac{\mathrm{SNR}_T}{\mathrm{SNR}_X}=\sqrt2\,\frac{|W_{\rm Re}|}{|W_H|} .
\label{eq:ratioSNR}
\end{eqnarray}
\end{subequations}
Since $a$ is taken real, this ratio equals $\sqrt2\,|\cos\chi|$, where
$\chi$ is the argument of the postselection amplitude $a+b$. It equals
$\sqrt2$ for a linear polarisation ($\mathrm{Im}(b)=0$ implies $\chi=0$),
for which the spectral channel contributes noise only; $1$ for the states
whose postselection amplitude has real and imaginary parts of equal
modulus ($\chi=\pm45^\circ$); and $0$ for the states with
$\mathrm{Re}(b)=-a$ ($\chi=\pm90^\circ$), for which the signal is carried
by the spectral centroid alone. The appropriate choice of estimator
therefore depends on the family of polarisation states considered, as
shown in section~\ref{sec:resolution}.

\subsection{The three floors on the postselection amplitude}
\label{sec:floor}

The uncertainty on the reconstruction presented in section~\ref{sec:stat}
is evaluated as the input state $|\psi\rangle$ approaches $|A\rangle$, in
order to determine the minimum value $\varepsilon_{\rm min}$ of the
postselection amplitude that fixes the radius of the forbidden region in
equation~\eqref{eq:cap}. Near $|A\rangle$, $a$ tends to $1/\sqrt2$, and the
modulus of the weak value defined in equation~\eqref{eq:Wmod} and the
uncertainty $\delta X$ defined in equation~\eqref{eq:dXeps} become
\begin{subequations}
\begin{eqnarray}
|W_H|=\frac{1}{2\varepsilon},
\label{eq:Wnear}\\
|\delta X|=\frac{\sqrt2}{x\,\varepsilon\sqrt{N_{\rm total}}},
\label{eq:dXnear}
\end{eqnarray}
\end{subequations}
so their ratio is independent of $\varepsilon$ and equals
$2\sqrt2/(x\sqrt{N_{\rm total}})$. $\mathrm{SNR}_T$ is likewise
independent of $\varepsilon$~\eqref{eq:SNRT}, and $\mathrm{SNR}_X$ does
not depend on it~\eqref{eq:SNRX}. As the input state approaches
$|A\rangle$ there is therefore no improvement in the relative uncertainty
of the reconstruction, since the divergence of the two centroids $T$ and
$\Omega$ is exactly compensated by the reduction of the number of
postselected photons $N$ defined in equation~\eqref{eq:Nps}, in agreement
with the metrological analyses of weak-value amplification
\cite{KneeGauger2014,FerrieCombes2014}. This absence of improvement
concerns the non-normalised coefficients $\psi_H$ and $\psi_V$ of
equation~\eqref{eq:direct}, but not the reconstructed state itself. For
the reconstructed state, the angular resolution presented in
section~\ref{sec:resolution} improves without limit near $|A\rangle$: that
section parametrises a path on the Poincar\'e sphere by an angle $\varphi$
and establishes that the sensitivity $|dW_H/d\varphi|$ diverges as
$1/\varepsilon^2$, whereas the modulus of the weak value $|W_H|$ and its
uncertainty diverge only as $1/\varepsilon$. The minimum value
$\varepsilon_{\rm min}$ therefore does not come from the uncertainty on
the two centroids, and its origin is established in the remainder of this
section.

The first of the three floors is caused by the dark counts of the
detector. Over an acquisition time $t_{\rm acq}$, a detector with an
equivalent dark-count rate $\Gamma$ records a number of dark counts
$N_{\rm dark}=\Gamma t_{\rm acq}$. These counts are distributed uniformly
over the detection window, which is centred on the undisplaced pointer
component, so they pull the measured mean arrival time $T$ towards that
centre and introduce a systematic shift $\Delta T=-rT$ to first order in
the ratio $r=N_{\rm dark}/N$. The relative bias on the mean arrival time
$T$ is therefore $-r$ and does not depend on $\tau$, since $T=\tau W_{\rm
Re}$ from equation~\eqref{eq:Texact}. This first-order treatment assumes
$r$ to be small, hence that the number of postselected counts exceeds the
number of dark counts, $N\ge N_{\rm dark}$, which by
equation~\eqref{eq:Nps}, with $\mathcal{N}=\varepsilon$ in the weak
regime, gives $\varepsilon\ge\varepsilon_{\rm det}$, with
\begin{equation}
\varepsilon_{\rm det}=\sqrt{\frac{\Gamma t_{\rm acq}}{N_{\rm total}}} .
\label{eq:eps2}
\end{equation}
Below $\varepsilon_{\rm det}$ the dark counts exceed the postselected counts. The bias
$\Delta T$ must in addition stay below the statistical error $\delta T$ of
equation~\eqref{eq:dT}, which, with the mean arrival time of
equation~\eqref{eq:Texact}, requires $N_{\rm dark}\le\sqrt N/(x\,|W_{\rm
Re}|)$. This second condition sets a tolerance on $\Delta T$ rather than a
condition of validity, and it depends on the coupling and on the state
$|\psi\rangle$, so only the condition of equation~\eqref{eq:eps2} enters
the definition of the forbidden region of equation~\eqref{eq:epsmin}.
Since $r$ grows as $1/\varepsilon^2$ as the state $|\psi\rangle$
approaches $|A\rangle$, whereas the relative uncertainty of
equation~\eqref{eq:rel} stays constant, the dark counts limit the
reconstruction near $|A\rangle$ through the condition introduced in
equation~\eqref{eq:eps2}. This limit cannot be removed by accumulating
more events, since the bias $\Delta T$ is systematic, but time gating
would reduce $N_{\rm dark}$ in proportion to the duty cycle. The floor
$\varepsilon_{\rm det}$ depends on the detector through $\Gamma$ alone,
whereas the statistical model of section~\ref{sec:stat} holds equally for
photon counting and for linear detection. The detection principle
therefore does not enter the reconstruction \cite{Jordan2014}.

The second floor follows from that same statistical model, where the
uncertainties on the two centroids are those of sample means. A sample
mean follows a normal distribution only above a minimum number $N_{\rm
min}$ of postselected events, typically of the order of ten, so with the
number of postselected events of equation~\eqref{eq:Nps} the condition
$N\ge N_{\rm min}$ reads $\varepsilon\ge\varepsilon_{\rm stat}$, with
\begin{equation}
\varepsilon_{\rm stat}=\sqrt{\frac{N_{\rm min}}{N_{\rm total}}} .
\label{eq:eps1}
\end{equation}

The third floor comes from the departure of the postselection probability
from its zero-coupling value $\varepsilon^2$, an equality that is a
condition of validity of the weak-value description \cite{Kofman2012}. The
exact expression of equation~\eqref{eq:N2} separates that probability into
this value and a correction,
$\mathcal{N}^2=\varepsilon^2-2\,\mathrm{Re}(A\bar B)(1-q)$, where $A$ and
$B$ are the pointer amplitudes of equation~\eqref{eq:Ft}. Near
$|A\rangle$, these two amplitudes tend to $1/2$ and $-1/2$, so the
postselection probability becomes $\mathcal{N}^2=\varepsilon^2+(1-q)/2$.
The overlap defined in equation~\eqref{eq:q} departs from unity by
$x^2/8$, so the correction exceeds $\varepsilon^2$ below a postselection
amplitude
\begin{equation}
\varepsilon_{\rm weak}=\frac{x}{4},
\label{eq:eps3}
\end{equation}
a floor that does not depend on $N_{\rm total}$. The same departure of $q$
from unity shifts the measured quantity $X$ from the weak value $W_H$ by
$\Delta X=X-W_H$, which section~\ref{sec:optimal} determines to first
order in $1-q$, and for linear polarisation states near $|A\rangle$ that
shift is a fraction of the weak value given by
\begin{equation}
\frac{|\Delta X|}{|W_H|}=\left(\frac{\varepsilon_{\rm weak}}{\varepsilon}\right)^{2} .
\label{eq:bias3}
\end{equation}
As $|\psi\rangle$ runs along the family of linear polarisation states, $X$
reaches a maximum at $\varepsilon_{\rm weak}$, up to a relative correction
of order $x^2$, so two states lying on either side of that amplitude give
the same mean arrival time $T$ and the same mean frequency shift $\Omega$,
as established in section~\ref{sec:fold}. The measurement of these two
centroids therefore no longer determines the state below $\varepsilon_{\rm
weak}$~\cite{Piacentini2018}.

\begin{figure}[htbp]
\centering
\includegraphics[width=0.75\linewidth]{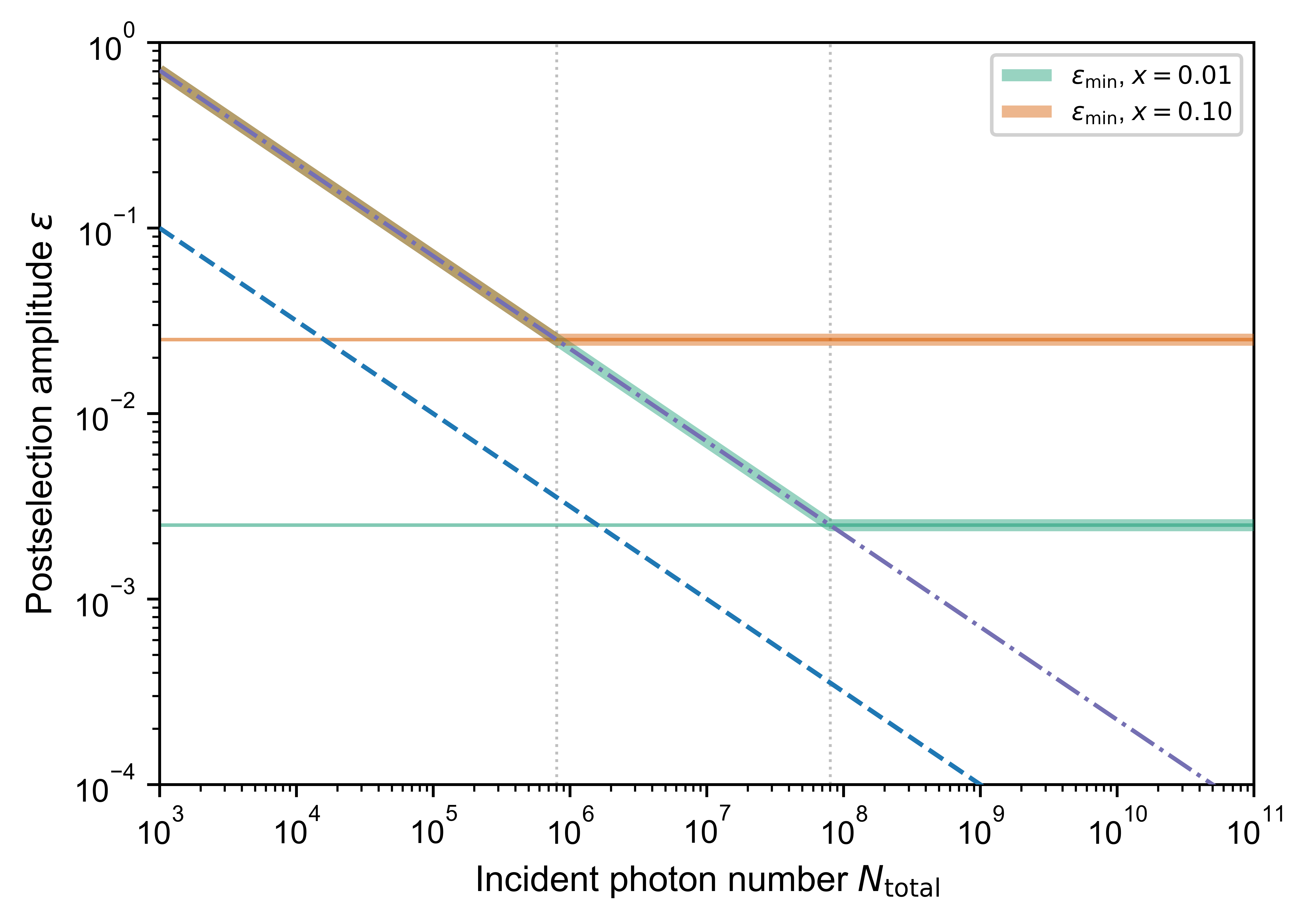}
\caption{\label{fig:floors}Postselection amplitude below which the
reconstruction fails, $\varepsilon_{\rm min}$ (thick bands, green
$x=0.01$, orange $x=0.10$), against the incident photon number. Each band
follows the detector floor $\varepsilon_{\rm det}$ (dash-dotted) and falls
as $N_{\rm total}^{-1/2}$ until it meets the weak-regime floor
$\varepsilon_{\rm weak}=x/4$ (thin horizontals) at the crossover marked by
the dotted verticals, beyond which $\varepsilon_{\rm min}$ no longer
depends on $N_{\rm total}$. The excluded cap has geodesic radius
$\alpha_{\rm min}=2\arcsin(\varepsilon_{\rm min})$. At
$N_{\rm total}=5\times10^{6}$ the weaker coupling is detector-limited,
$\varepsilon_{\rm min}=1.0\times10^{-2}$ and
$\alpha_{\rm min}=1.1^\circ$, while the stronger is saturated,
$\varepsilon_{\rm min}=2.5\times10^{-2}$ and
$\alpha_{\rm min}=2.9^\circ$. The statistical floor
$\varepsilon_{\rm stat}$ (dashed) is never the largest of the three.
Parameters $N_{\rm min}=10$, $\Gamma=100$~s$^{-1}$, $t_{\rm acq}=5$~s.}
\end{figure}

These three floors set a lower bound on the postselection amplitude
$\varepsilon$, so the smallest admissible value is the largest of them,
\begin{equation}
\varepsilon_{\rm min}=\max\left(\varepsilon_{\rm stat},
\varepsilon_{\rm det},\varepsilon_{\rm weak}\right),
\label{eq:epsmin}
\end{equation}
as shown in figure~\ref{fig:floors}. Equations~\eqref{eq:eps1}
and~\eqref{eq:eps2} show that the ratio $\varepsilon_{\rm
stat}/\varepsilon_{\rm det} =\sqrt{N_{\rm min}/\Gamma t_{\rm acq}}$ is
below unity as soon as the number of dark counts exceeds $N_{\rm min}$, so
the statistical floor can be removed from equation~\eqref{eq:epsmin} and
$\varepsilon_{\rm min}$ is the larger of $\varepsilon_{\rm det}$ and
$\varepsilon_{\rm weak}$. These two floors cross at
\begin{equation}
N_{\rm total}^{*}=\frac{16\,\Gamma t_{\rm acq}}{x^2},
\label{eq:cross}
\end{equation}
so $\varepsilon_{\rm min}$ decreases as $N_{\rm total}^{-1/2}$ below that
value and equals $x/4$ above it. With $\Gamma=100$~s$^{-1}$, $t_{\rm
acq}=5$~s and $N_{\rm total}=5\times10^{6}$, $N_{\rm total}$ lies below
the crossover for $x=0.01$ and above it for $x=0.10$, as shown in
table~\ref{tab:floors}. Since $\varepsilon_{\rm min}$ is small, an
expansion of equation~\eqref{eq:cap} gives the radius of the forbidden
region as $\alpha_{\rm min}=2\varepsilon_{\rm min}$, and both radii in the
table stay below $3^\circ$. Raising the coupling from $x=0.01$ to $x=0.10$
multiplies the signal-to-noise ratio $\mathrm{SNR}_X$ by ten and
$\alpha_{\rm min}$ by $2.5$ only, the two factors differing because the
radius of the excluded region at the weaker coupling is set by the
detector floor $\varepsilon_{\rm det}$ and not by the weak-regime floor
$\varepsilon_{\rm weak}$.

Weak-value amplification \cite{Kofman2012} uses the same divergence for a
different purpose, since it seeks the largest pointer displacement,
whereas the reconstruction requires the three floors of
equation~\eqref{eq:epsmin} to hold. Section~\ref{sec:fold} establishes
that an analytical inversion of the relation between $X$ and the state at
non-zero coupling is possible, so $\varepsilon_{\rm weak}$ no longer
bounds the reconstruction but the two-centroid readout, that is the
amplitude below which the two centroids are no longer sufficient to
determine the state.

\begin{table}[htpb]
\centering
\caption{The three floors on the postselection amplitude and the resulting
exclusion cap, for the two couplings used throughout.
$\varepsilon_{\rm min}$ is the largest of the three floors of
equation~\eqref{eq:epsmin}, and $\alpha_{\rm min}=2\varepsilon_{\rm min}$
is the geodesic half-angle of the excluded cap. Parameters
$N_{\rm min}=10$, $\Gamma=100$~s$^{-1}$, $t_{\rm acq}=5$~s, incident flux
$10^{6}$~s$^{-1}$, hence $N_{\rm total}=5\times10^{6}$.}
\label{tab:floors}
\vspace{6pt}
\begin{tabular}{c c c c c c c}
\hline
\multicolumn{1}{c}{$x$} & \multicolumn{1}{c}{$\varepsilon_{\rm stat}$}
 & \multicolumn{1}{c}{$\varepsilon_{\rm det}$}
 & \multicolumn{1}{c}{$\varepsilon_{\rm weak}$}
 & \multicolumn{1}{c}{$N_{\rm total}^{*}$}
 & \multicolumn{1}{c}{$\varepsilon_{\rm min}$}
 & \multicolumn{1}{c}{$\alpha_{\rm min}$}\\
 & & & & & & \multicolumn{1}{c}{(deg)}\\
\hline
$0.01$ & $1.4\times10^{-3}$ & $1.0\times10^{-2}$ & $2.5\times10^{-3}$
 & $8.0\times10^{7}$ & $1.0\times10^{-2}$ & $1.1$\\
$0.10$ & $1.4\times10^{-3}$ & $1.0\times10^{-2}$ & $2.5\times10^{-2}$
 & $8.0\times10^{5}$ & $2.5\times10^{-2}$ & $2.9$\\
\hline
\end{tabular}
\end{table}

\subsection{Information beyond the centroids}
\label{sec:profile}

Since the two centroids are two numbers extracted from the postselected
profile $|F(t)|^2$, whose amplitude is defined in equation~\eqref{eq:Ft},
the Fisher information that this profile carries on the parameter
describing the state is compared with the information that these centroids
carry on that parameter. As $x$ tends to zero, the profile is a Gaussian
displaced by $\tau W_{\rm Re}$, so its centroid alone contains all the
information that this profile carries on the state. At non-zero coupling,
the whole profile carries more information on the state than the centroid
alone \cite{KneeGauger2014}.

Let $p(t)$ be the postselected profile normalised so that its integral
equals one, and let $\theta$ be the angle characterising the family of
linear polarisation states
$|\psi\rangle=\cos\theta\,|H\rangle+\sin\theta\,|V\rangle$. The Fisher
information carried by $p$ on $\theta$ and that carried by the centroid
alone are
\begin{equation}
I_{\rm prof}=\int\frac{1}{p}\left(\frac{\partial p}{\partial\theta}\right)^{2}dt,
\qquad
I_{\rm cent}=\frac{1}{\sigma^{2}}\left(\frac{dT}{d\theta}\right)^{2}
\label{eq:Iprof}
\end{equation}
per postselected photon, and the variance of any unbiased estimator of
$\theta$ built on either of these two readings is at least the inverse of
the corresponding information, divided by the number of photons. $I_{\rm
cent}$ has an explicit form obtained from the arrival time defined in
equation~\eqref{eq:Texact}, whereas $I_{\rm prof}$ is computed numerically
from the profile of equation~\eqref{eq:Ft}. Since only a fraction
$\mathcal{N}^2$ of the incident photons is postselected, and since the
Fisher information adds over independent photons, the information per
incident photon is obtained by multiplying $I_{\rm prof}$ and $I_{\rm
cent}$ by $\mathcal{N}^2$, which gives $\mathcal{N}^2I_{\rm prof}$ for the
profile and $\mathcal{N}^2I_{\rm cent}$ for the centroid. Both
informations are listed in table~\ref{tab:fisher}.

When the radius of the forbidden region is set by the detector floor
$\varepsilon_{\rm det}$, the two informations are nearly equal, and for a
coupling $x=0.01$ they differ by only two per cent for a state lying
$4^\circ$ from $|A\rangle$ ($\varepsilon=3.49\times10^{-2}$). When that
radius is set by the weak-regime floor, the boundary of the region lies
where $X$ reaches its maximum and $\mathcal{N}^2I_{\rm cent}$ vanishes,
whereas $\mathcal{N}^2I_{\rm prof}$ has a finite value. The two
informations already differ before that maximum amplitude is reached,
since for $x=0.4$ the corresponding floor is $\varepsilon_{\rm
weak}=0.100$ and $\mathcal{N}^2I_{\rm cent}=4.04\times10^{-2}$ against
$2.26$ for $\mathcal{N}^2I_{\rm prof}$ for a state lying $10^\circ$ from
$|A\rangle$ ($\varepsilon=8.72\times10^{-2}$).

\begin{table}[htpb]
\centering
\caption{Fisher information on the polarisation angle $\theta$ of the
linear family $|\psi\rangle=\cos\theta\,|H\rangle+\sin\theta\,|V\rangle$,
for postselection on $|D\rangle$, carried by the centroid alone and by the
whole profile, per incident photon. Both informations are dimensionless
and depend on the coupling only through $x=\tau/\sigma$, so no separate
value of $\tau$ or $\sigma$ is required. The postselection amplitude is
$\varepsilon=\sin(\alpha/2)$, with $\alpha$ the geodesic angle between
$|\psi\rangle$ and $|A\rangle$; the eight rows sample
$\alpha=0.4^\circ$, $1^\circ$, $4^\circ$, $10^\circ$ and $20^\circ$, on
either side of $\varepsilon_{\rm weak}=x/4$ for each coupling.
$I_{\rm cent}$ follows from equation~\eqref{eq:Iprof} in explicit form,
$I_{\rm prof}$ is evaluated numerically.}
\label{tab:fisher}
\vspace{6pt}
\begin{tabular}{c c c c c}
\hline
\multicolumn{1}{c}{$x$} & \multicolumn{1}{c}{$\varepsilon$}
 & \multicolumn{1}{c}{$\mathcal{N}^2$}
 & \multicolumn{1}{c}{$\mathcal{N}^2I_{\rm cent}$}
 & \multicolumn{1}{c}{$\mathcal{N}^2I_{\rm prof}$}\\
\hline
$0.01$ & $3.49\times10^{-2}$ & $1.22\times10^{-3}$ & $2.00\times10^{-2}$ & $2.04\times10^{-2}$\\
$0.01$ & $3.49\times10^{-3}$ & $1.84\times10^{-5}$ & $1.40\times10^{-1}$ & $1.36$\\[4pt]
$0.10$ & $8.72\times10^{-2}$ & $8.22\times10^{-3}$ & $2.19\times10^{-1}$ & $3.04\times10^{-1}$\\
$0.10$ & $3.49\times10^{-2}$ & $1.84\times10^{-3}$ & $1.41\times10^{-1}$ & $1.36$\\
$0.10$ & $3.49\times10^{-3}$ & $6.37\times10^{-4}$ & $3.63$ & $3.92$\\[4pt]
$0.40$ & $1.74\times10^{-1}$ & $3.96\times10^{-2}$ & $2.68\times10^{-1}$ & $9.91\times10^{-1}$\\
$0.40$ & $8.72\times10^{-2}$ & $1.74\times10^{-2}$ & $4.04\times10^{-2}$ & $2.26$\\
$0.40$ & $8.73\times10^{-3}$ & $9.98\times10^{-3}$ & $3.89$ & $3.93$\\
\hline
\end{tabular}
\end{table}

The Fisher information of the profile $\mathcal{N}^2I_{\rm prof}$ does not
vanish where that of the centroid vanishes, so the minimum variance of an
estimator based on the profile stays finite there, whereas that of an
estimator based on the two centroids diverges. The weak-regime floor
$\varepsilon_{\rm weak}$ is therefore relevant to $\varepsilon_{\rm min}$
only when the reconstruction is based on the two centroids. A
reconstruction method based on the whole profile has only two floors,
$\varepsilon_{\rm stat}$ and $\varepsilon_{\rm det}$, that is
\begin{equation}
\varepsilon_{\rm min}^{\rm(prof)}=\max\left(\varepsilon_{\rm stat},
\varepsilon_{\rm det}\right),
\label{eq:epsminprof}
\end{equation}
both of which decrease as $N_{\rm total}^{-1/2}$. The state is then
reconstructed by comparing the measured profile with computed theoretical
profiles, and no longer through the inversion presented in
section~\ref{sec:exact}. Section~\ref{sec:resolution} is based on the two
centroids, so the three floors bound the postselection amplitude there.

%% file: section4_angular_resolution.tex
\section{Angular resolution along an arbitrary path}
\label{sec:resolution}

A test of a reconstruction method characterises a continuum of states rather
than a single state, since rotating a waveplate makes the input state
$|\psi(\varphi)\rangle$ follow a path on the Poincar\'e sphere. The
reconstruction is then performed on a discrete set of states along that path,
and the step $\Delta\varphi$ between two settings of the angle $\varphi$ is the
quantity whose minimal value is determined in this section. Two settings
separated by less than that minimum return values of $X$ that the measurement
cannot separate, so this minimum is also the smallest change of
$|\psi(\varphi)\rangle$ that the method can resolve. Throughout this section the
configuration assumes postselection on $|\varsigma\rangle=|D\rangle$, and the
two coupling values $x=0.01$ and $x=0.10$ introduced in
section~\ref{sec:errors} are used as illustrations.

\subsection{The weak value as a stereographic projection}
\label{sec:proj}

Determining the minimal value of $\Delta\varphi$ requires the weak value written
in terms of the Stokes parameters. With the Stokes parameters introduced in
section~\ref{sec:cap}, the weak value of a state $|\psi(\varphi)\rangle$ of
Stokes vector
$\mathbf{S}(\varphi)=(S_1(\varphi),S_2(\varphi),S_3(\varphi))$ takes the form
\begin{equation}
W_H=\frac{1}{2}+\frac{\zeta}{2},
\qquad
\zeta=\frac{S_1-iS_3}{1+S_2},
\label{eq:Wstokes}
\end{equation}
whose denominator is twice the postselection probability introduced in
equation~\eqref{eq:N2S2}. That denominator vanishes when
$|\psi\rangle=|A\rangle$, where $W_H$ diverges. The bijection that maps any
point of the Poincar\'e unit sphere other than $|A\rangle$, corresponding to the
state $|\psi\rangle$ represented by the vector $\mathbf{S}$, to
$(S_1,S_3)/(1+S_2)$ is a stereographic projection from the point associated with
the polarisation $|A\rangle$ onto the plane $S_2=0$ \cite{Kobayashi2014}, so
that $\zeta$ is the complex coordinate of the projected point, from which the
weak value follows through equation~\eqref{eq:Wstokes}.

\subsection{The resolution criterion}
\label{sec:criterion}

The resolution criterion follows from the noise carried by the two centroids.
The definition of $X$ in equation~\eqref{eq:X} gives the uncertainty on each of
its two parts from the centroid uncertainties of equations~\eqref{eq:dT}
and~\eqref{eq:dOm}, namely $\delta T/\tau$ for the real part and
$(2\sigma^2/\tau)\,\delta\Omega$ for the imaginary part. Since
$\sigma/\tau=1/x$, both reduce to the same value
\begin{equation}
s=\frac{1}{x\sqrt{N}},
\label{eq:s}
\end{equation}
so that the pointer width cancels and the noise is isotropic in the complex
plane of $X$. The arrival times are distributed about the temporal centroid,
that is the mean arrival time $T=\tau W_{\mathrm{Re}}$ (see
equation~\eqref{eq:Texact}), with width $\sigma$, whereas the frequency shifts
are distributed about the spectral centroid
$\Omega=(\tau/2\sigma^2)\,W_{\mathrm{Im}}$ (see equation~\eqref{eq:Omexact}),
with width $1/(2\sigma)$. Both widths are those of the pointer alone, since
the temporal response of the detector is taken to be small compared with
$\sigma$ and its spectral resolution small compared with $1/\sigma$. Each
centroid provides a Fisher information equal to
$Nx^2$ times the squared derivative of the relevant part, $W_{\mathrm{Re}}$ or
$W_{\mathrm{Im}}$, of the weak value, where $N$ is the number of postselected
photons (see equation~\eqref{eq:Nps}). Since the two distributions are measured
on distinct sets of photons, they are independent and their Fisher informations
add \cite{Hofmann2011,Dressel2014}:
\begin{equation}
I(\varphi)=Nx^2\left|\frac{dW_H}{d\varphi}\right|^2 .
\label{eq:fisher}
\end{equation}

The Cram\'er--Rao bound then fixes the minimal step,
\begin{equation}
\Delta\varphi_{\min}=\frac{1}{x\sqrt{N}\,\left|dW_H/d\varphi\right|}.
\label{eq:crit}
\end{equation}
For a family of linear states ($S_3=0$), the weak value is real, only the mean
arrival time $T$ is measured, and $|dW_H/d\varphi|$ becomes
$|dW_{\mathrm{Re}}/d\varphi|$. This bound applies to the estimator used here,
which extracts two centroids from the data. An estimator using the complete
postselected profile is not bounded by it, and section~\ref{sec:profile}
compares the Fisher informations of these two extraction methods. As $\varphi$
varies, $X$ moves in a known direction of the complex plane, and only the noise
component along that direction contributes to the uncertainty on $\varphi$. The
noise entering the bound is therefore $s$ (see equation~\eqref{eq:s}) and not
the modulus $|\delta X|$ (see equation~\eqref{eq:dX}).

For an input state $|\psi(\varphi_1,\varphi_2)\rangle$ characterised by two
parameters, the states reached no longer form a path but a region on the surface
of the Poincar\'e sphere. This case would follow the same reasoning, with two
derivatives instead of one, which would give the Fisher matrix
\begin{equation}
I_{ij}=x^2N\,\mathrm{Re}\!\left[
\left(\frac{dW_H}{d\varphi_i}\right)^{\!*}\frac{dW_H}{d\varphi_j}\right].
\label{eq:fishermat}
\end{equation}
Its inverse is the covariance ellipse of the two estimates, and the matrix
reduces to equation~\eqref{eq:fisher} when the two indices coincide.

\subsection{Sensitivity along an arbitrary path}
\label{sec:path}

Evaluating the criterion of equation~\eqref{eq:crit} for a given path on the
Poincar\'e sphere requires $dW_H/d\varphi$. Equation~\eqref{eq:Wstokes} is
therefore differentiated and the real and imaginary parts of the numerator are
separated, which gives
\begin{equation}
\frac{dW_H}{d\varphi}=\frac{\mathcal{D}}{2(1+S_2)^2},
\qquad
\mathcal{D}=\left[S_1'(1+S_2)-S_1S_2'\right]
-i\left[S_3'(1+S_2)-S_3S_2'\right],
\label{eq:Dpath}
\end{equation}
where the prime denotes the derivative with respect to $\varphi$. Inserting this
derivative into the criterion and using $\mathcal{N}=\sqrt{(1+S_2)/2}$ (see
equation~\eqref{eq:N2S2}), the minimal step for any path is
\begin{equation}
\Delta\varphi_{\min}
=\frac{2\sqrt{2}\,(1+S_2)^{3/2}}{x\sqrt{N_{\mathrm{total}}}\,|\mathcal{D}|}.
\label{eq:critgen}
\end{equation}
The minimal step is thus obtained from the path itself: $S_2$ entirely
determines the postselection probability, and $\mathcal{D}$ determines how much
the weak value changes as the state moves along the path, that is, the
sensitivity of the weak value.

\subsection{Two paths}
\label{sec:examples}

The general expression is now applied to two families of states
$|\psi\rangle=a|H\rangle+b|V\rangle$, both generated by rotating a single
waveplate and parametrised by a single angle $\theta$. For the first path,
corresponding to the linear polarisation states, $\theta$ is directly the angle
of the polarisation state, that is $a=\cos\theta$ and $b=\sin\theta$, or in the
Stokes representation $S_1=\cos 2\theta$, $S_2=\sin 2\theta$ and $S_3=0$, with
which the postselection probability of equation~\eqref{eq:N2lin} is associated.
Since the Stokes parameters depend on $2\theta$, the interval $[0,\pi)$ is
sufficient to describe the whole great circle. On this path $\mathcal{D}$
reduces to $-2(1+S_2)$. Consequently $dW_H/d\theta$ equals
$-1/(2\mathcal{N}^2)$, which gives a minimal step
\begin{equation}
\Delta\theta_{\min}=\frac{2\mathcal{N}}{x\sqrt{N_{\mathrm{total}}}}.
\label{eq:steplin}
\end{equation}
The weak value is real along this path, so $W_{\mathrm{Im}}=0$ and $\Omega=0$
for all its states, and determining the mean arrival time $T$ is sufficient. The
step is proportional to $\mathcal{N}$, so it is coarsest when $|\psi\rangle$
coincides with $|D\rangle$, where the postselection probability is maximal and
the sensitivity of the weak value minimal, and it decreases as $|\psi\rangle$
approaches $|A\rangle$. In the weak regime $\mathcal{N}=\varepsilon$, so the
sensitivity $|dW_H/d\theta|=1/(2\varepsilon^2)$ increases as the postselection
amplitude tends to zero. This decrease stops at the boundary of the excluded
region defined in section~\ref{sec:cap}, when
$\mathcal{N}=\varepsilon_{\min}$.

The second path, corresponding to the states $|\psi\rangle$ lying on the great
circle $S_2=0$ and passing through the states $|H\rangle$, $|R\rangle$,
$|V\rangle$ and $|L\rangle$, is parametrised by $a=\cos\theta$ and
$b=i\sin\theta$, or in the Stokes representation $S_1=\cos 2\theta$, $S_2=0$ and
$S_3=\sin 2\theta$, with a postselection probability constant and equal to
$1/2$. On this path $\mathcal{D}$ reduces to $-2\sin 2\theta-2i\cos 2\theta$.
Consequently $dW_H/d\theta$ equals $-\sin 2\theta-i\cos 2\theta$, which gives a
minimal step
\begin{equation}
\Delta\theta_{\min}=\frac{\sqrt{2}}{x\sqrt{N_{\mathrm{total}}}},
\label{eq:stepmer}
\end{equation}
independent of $\theta$ since $|dW_H/d\theta|=1$. When
$|\psi\rangle=|H\rangle$ or $|V\rangle$, $dW_H/d\theta$ is purely imaginary, so
the mean frequency shift alone carries the whole variation. A scan at constant
step along this great circle $S_2=0$ therefore requires both observables,
temporal and spectral, since the mean arrival time alone does not separate two
states neighbouring $|H\rangle$ or $|V\rangle$.

Three great circles illustrate the general case. On $S_3=0$ the postselection
probability varies along the path and $\mathcal{D}$ is real, so the mean arrival
time carries the whole variation. On $S_2=0$ that probability is constant and
both observables are required. On $S_1=0$, which passes through $|D\rangle$,
$|R\rangle$, $|A\rangle$ and $|L\rangle$, $\mathcal{D}$ is purely imaginary, so
the mean frequency shift carries the whole variation. This third path is the
counterpart of the first and, like it, crosses the excluded region defined in
section~\ref{sec:cap} when $|\psi\rangle=|A\rangle$. An arbitrary path combines
these behaviours, and $\mathcal{D}$ of equation~\eqref{eq:Dpath} gives them
without further calculation.

\subsection{The step in numbers}
\label{sec:numbers}

\begin{figure}[htpb]
\centering
\includegraphics[width=0.75\linewidth]{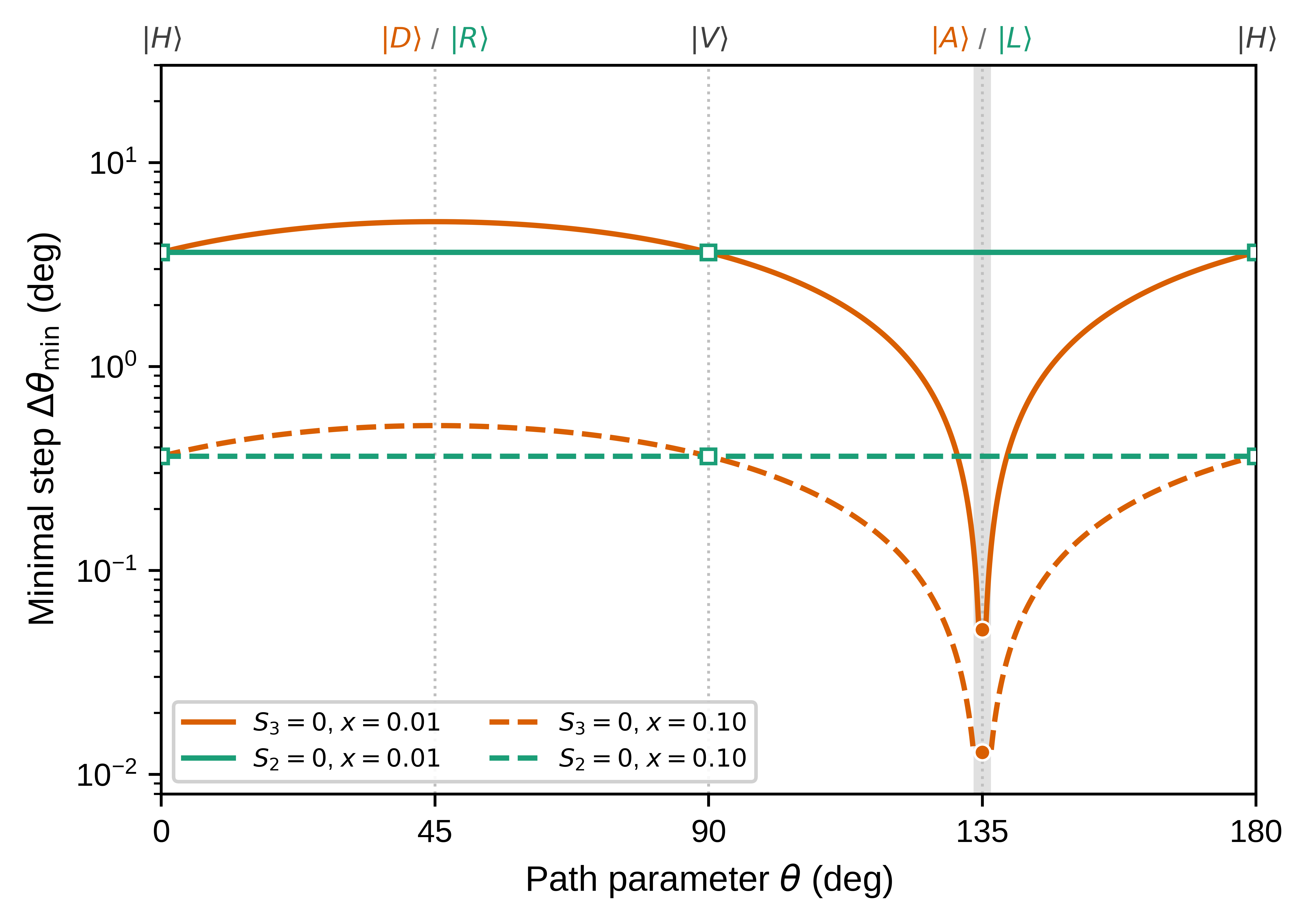}
\caption{Minimal step $\Delta\theta_{\min}$ along the two paths, against the
path parameter $\theta$, at $N_{\mathrm{total}}=5\times10^{6}$ incident photons
per setting. Orange: the great circle $S_3=0$, of linear polarisation states.
Green: the great circle $S_2=0$. Solid lines: $x=0.01$. Dashed lines:
$x=0.10$. The upper labels name the state reached at that value of $\theta$ on
each path, in the colour of the path concerned. On $S_2=0$ the step is
independent of $\theta$, and the open squares mark $|H\rangle$ and $|V\rangle$,
where the derivative is purely imaginary and the mean arrival time alone does
not separate two neighbouring settings. On $S_3=0$ the step decreases towards
$|A\rangle$, the shaded bands being the excluded region of section~\ref{sec:cap}
at each coupling; the filled circles mark the smallest accessible step,
$0.051^{\circ}$ at $x=0.01$ and $0.013^{\circ}$ at $x=0.10$.}
\label{fig:steps}
\end{figure}

With a number of incident photons $N_{\mathrm{total}}=5\times10^{6}$ per state
$|\psi(\theta)\rangle$, as in section~\ref{sec:floor}, the common factor
$x\sqrt{N_{\mathrm{total}}}$ equals $22.4$ for a coupling $x=0.01$ and $224$ for
$x=0.10$. Figure~\ref{fig:steps} shows the minimal step as a function of
$\theta$ for the two paths and at the two couplings. On the path along the great
circle $S_2=0$ the minimal step is constant, whereas on the path corresponding
to the equator ($S_3=0$) it reaches its maximal value when
$|\psi\rangle=|D\rangle$ and decreases as $|\psi\rangle$ tends towards
$|A\rangle$, down to the boundary of the excluded region.

\begin{table}[htpb]
\centering
\caption{Minimal step $\Delta\theta_{\min}$ at three states of the equator
($S_3=0$) and along the great circle $S_2=0$, for
$N_{\mathrm{total}}=5\times10^{6}$ incident photons per state and the two
couplings of section~\ref{sec:errors}. The second column gives the observables
that have to be recorded, the mean arrival time alone being sufficient wherever
the weak value is real.}
\label{tab:steps}
\vspace{6pt}
\small
\setlength{\tabcolsep}{5pt}
\begin{tabular}{@{}p{0.34\linewidth}p{0.30\linewidth}cc@{}}
\hline
\multicolumn{1}{c}{path / state} & \multicolumn{1}{c}{observables}
& \multicolumn{1}{c}{$x=0.01$} & \multicolumn{1}{c}{$x=0.10$}\\
\hline
equator, at $|H\rangle$ & mean arrival time & $3.62^{\circ}$ & $0.362^{\circ}$\\
equator, at $|D\rangle$ & mean arrival time & $5.12^{\circ}$ & $0.512^{\circ}$\\
equator, at the boundary of the excluded region & mean arrival time
& $0.051^{\circ}$ & $0.013^{\circ}$\\
great circle $S_2=0$, any state & mean arrival time $+$ frequency shift
& $3.62^{\circ}$ & $0.362^{\circ}$\\
\hline
\end{tabular}
\end{table}

Every entry of table~\ref{tab:steps} varies as the inverse of
$x\sqrt{N_{\mathrm{total}}}$, so that passing from $x=0.01$ to $x=0.10$ divides
it by ten, except at the boundary of the excluded region, where
$\mathcal{N}=\varepsilon_{\min}$ depends on the coupling as well and the entry
is divided by four only. Dividing the range of $\theta$, which runs from $0$ to
$180^{\circ}$, by the coarsest step gives the number of values of $\theta$
needed to cover a whole path: about $35$ values at $x=0.01$ and about $350$ at
$x=0.10$. For a situation in which the delay $\tau$ is fixed by the experimental
set-up, the coupling $x$ is modified by adjusting the pulse duration, and the
step is reduced by shortening it. The minimal step also decreases with the
number of incident photons $N_{\mathrm{total}}$, an increase of
$N_{\mathrm{total}}$ by a factor of $100$ giving a reduction of the step by a
factor of $10$. The coupling cannot, however, be increased indefinitely, since
the measured quantity $X$ departs from the weak value at order $x^2$, as
equation~\eqref{eq:Xq} shows, and the value that minimises the resulting error is established in
section~\ref{sec:optimal}.

%% file: section5_optimal_coupling.tex

\section{Optimal interaction strength}
\label{sec:optimal}

The measured quantity $X$ departs from the weak value $W$, and that
departure grows with the coupling, whereas the uncertainty on $X$ defined
in equation~\eqref{eq:s} decreases. This section determines the value of
the coupling that balances the two trends.

\subsection{Departure from the weak limit}
\label{sec:departure}

The bias on the reconstructed state follows from the difference between
$X$ and $W$, so that difference is evaluated first. Expanding $X(q)$ of
equation~\eqref{eq:Xq} to first order in $\eta=x^2/8$, which equals $1-q$
at that order, gives $X=W+\Delta X$ with
\begin{equation}
\Delta X=\eta\,\frac{R\left(2a^2-1\right)-iJ}{\left(1+2R\right)^2},
\label{eq:DeltaX}
\end{equation}
where $R$ and $J$ are the real and imaginary parts of $a\bar b$. This is
the departure of order $x^2$ mentioned in section~\ref{sec:resolution},
and the same quantity $1-q$ sets the floor $\varepsilon_{\rm weak}=x/4$
on the postselection amplitude in equation~\eqref{eq:eps3}, so the two
penalties for a stronger coupling are of the same order in the same
parameter. The corresponding difference for a general direct
reconstruction scheme is given in~\cite{Vallone2016}.

\subsection{Systematic error of the direct estimator}
\label{sec:syst}

The estimator that performs the inversion of the measured quantity $X$
introduced in section~\ref{sec:reconstruction}, without correction, is
called the direct estimator, and that inversion is valid only when $X$
equals $W$. At finite coupling, however, this quantity differs from $W$
by the deterministic departure $\Delta X$ of
equation~\eqref{eq:DeltaX}. The state is determined by the two
independent parameters $a$ and $\mathrm{Im}(b)$, $\mathrm{Re}(b)$
following from normalisation, and the systematic errors
$\delta a_{\rm syst}$ and $\delta\mathrm{Im}(b)_{\rm syst}$, which
constitute the bias of the direct estimator, are defined as the prepared
value minus the reconstructed value, $a-a_{\rm rec}$ and
$\mathrm{Im}(b)-\mathrm{Im}(b)_{\rm rec}$, both taken in the
representative that parametrises the path. This error comes entirely
from $\Delta X$ and would vanish if $X$ equalled $W$, and its propagation
through the inversion gives
\begin{equation}
\delta a_{\rm syst}=a-a_{\rm rec}
=-\mathrm{Re}\!\left[\left(P+iQ\right)\Delta X\right]
=\frac{x^2}{16}\,f_{\rm Re},
\qquad
\delta\mathrm{Im}(b)_{\rm syst}=\frac{x^2}{16}\,f_{\rm Im},
\label{eq:fRfI}
\end{equation}
where
\begin{equation}
P=\mathrm{Re}(b)+2a|b|^2,
\qquad
Q=\mathrm{Im}(b)\left(1-2a^2\right),
\label{eq:PQ}
\end{equation}
$f_{\rm Re}=a|b|^2\left(1-2a^2\right)/\mathcal{N}^2$ and
$f_{\rm Im}=\mathrm{Im}(b)\left(2a^4-a^2+2R\right)/\mathcal{N}^2$. A
stronger coupling therefore displaces both components of the state, and
the optimisation presented in section~\ref{sec:opt} applies to the pair.
The postselection probability $\mathcal{N}^2$ appears in the denominator
of both coefficients, so the systematic error diverges as $|\psi\rangle$
tends to $|A\rangle$, as $1/\varepsilon$ for $f_{\rm Re}$, whose
numerator vanishes at the same point, and as $1/\varepsilon^2$ for
$f_{\rm Im}$.

For a family of linear states ($S_3=0$) with $S_2=\sin 2\theta$,
$\mathrm{Im}(b)$ vanishes, so $f_{\rm Im}$ vanishes as well and
$f_{\rm Re}$ becomes
\begin{equation}
f(\theta)=-\frac{S_2\sin\theta\cos 2\theta}{1+S_2}.
\label{eq:ftheta}
\end{equation}
This coefficient vanishes at $|H\rangle$, $|D\rangle$ and $|V\rangle$,
where $X(q)$ of equation~\eqref{eq:Xq} equals $1$, $1/2$ and $0$ whatever
the value of $q$. For these three states the reconstruction is therefore
exact at any value of $x$, and the optimisation presented in
section~\ref{sec:opt} does not apply to them. On the great circle
$S_2=0$ of section~\ref{sec:examples}, the minimal step is independent of
$\theta$, and both the excluded region of section~\ref{sec:cap} and the
divergence of the systematic error remain in the neighbourhood of
$|A\rangle$, which does not belong to that path.

\subsection{Statistical error and optimal coupling}
\label{sec:opt}

The two couplings $x=0.01$ and $x=0.10$ of sections~\ref{sec:errors}
and~\ref{sec:resolution} illustrate the weak regime, in which the
reconstruction of section~\ref{sec:reconstruction} is exact. The present
section instead determines the coupling that minimises the error. The
statistical error propagates in the same way as the systematic error, the
deterministic departure $\Delta X$ being replaced by the isotropic noise
$s$ on the measured quantity $X$ (see equation~\eqref{eq:s}). That noise
originates in the finite number $N$ of postselected photons (see
equation~\eqref{eq:Nps}), which sets the uncertainties on the mean
arrival time $T$ and on the mean frequency shift $\Omega$. Propagating
these uncertainties through the definition of $X$ in
equation~\eqref{eq:X} gives an equal contribution $s$ to the real and
imaginary parts of $X$, so the noise is isotropic in the complex plane of
$X$. The two centroids are measured in different experimental
configurations, so their uncertainties are independent and their
variances add, and the statistical error is
\begin{equation}
\delta a_{\rm stat}=\frac{\sqrt K}{x\,\mathcal{N}\sqrt{N_{\rm total}}},
\qquad K=P^2+Q^2 .
\label{eq:dastat}
\end{equation}
For a family of linear states, $K$ equals $4b^2\mathcal{N}^4$, so
$\delta a_{\rm stat}$ equals $|b|$ times the minimal step defined in
equation~\eqref{eq:steplin}.

The mean squared error is the sum of the squared systematic error and the
variance,
\begin{equation}
E(x)=E_{\rm syst}+E_{\rm stat},
\qquad
E_{\rm syst}=\frac{f(\theta)^2x^4}{256},
\qquad
E_{\rm stat}=\frac{K}{x^2N} .
\label{eq:MSE}
\end{equation}
The first term increases with the coupling whereas the second decreases.
$E$ is therefore minimal at a specific value of the coupling, called
optimal, and equal to
\begin{equation}
x_{\rm opt}=\left(\frac{128\,K}{f(\theta)^2N}\right)^{1/6},
\label{eq:xopt}
\end{equation}
which varies as $N_{\rm total}^{-1/6}$. There,
\begin{equation}
E_{\rm stat}=2E_{\rm syst},
\label{eq:balance}
\end{equation}
and the residual error on $a$ is
\begin{equation}
\delta a_{\min}=0.546\left(\frac{|f(\theta)|\,K}{N}\right)^{1/3},
\label{eq:damin}
\end{equation}
which varies as $N_{\rm total}^{-1/3}$. Since $a=\cos\theta$ on this
family, the error on $a$ corresponds to the angular error
$\delta\theta(x)=\delta a(x)/|b|$, whose value at the optimal coupling,
$\delta\theta_{\min}$, equals $\sqrt{3/2}$ times the minimal step of
equation~\eqref{eq:steplin}, the factor arising from the residual bias.

$x_{\rm opt}$ depends on the state $|\psi\rangle$ only through the sixth
root of $K/f(\theta)^2$, so an estimate of that ratio in error by a
factor of two displaces $x_{\rm opt}$ by about twelve per cent. Writing
$\lambda=x/x_{\rm opt}$, the increase of the final error is
$\delta a(x)/\delta a_{\min}
=\sqrt{\left(\lambda^4+2/\lambda^2\right)/3}$, so a displacement of
twelve per cent increases the final error by less than three per cent.
For the two states of table~\ref{tab:opt}, which belong to the family of
linear states, $x_{\rm opt}$ varies by a factor of eight at a given
$N_{\rm total}$, whereas the angular error stays of the same order and
below three tenths of a degree at $10^6$ incident photons. This variation
grows as the state approaches $|A\rangle$, but the accessible range then
depends on the floor $\varepsilon_{\min}$ of section~\ref{sec:cap}, hence
on $\varepsilon_{\rm weak}$. The method thus reaches a comparable
precision over this whole family, at the cost of a pulse duration that
differs from one state to another.

\begin{table}[htpb]
\centering
\caption{Optimal coupling $x_{\rm opt}$ of equation~\eqref{eq:xopt}, the
expansion parameter $\eta$ it corresponds to, and the angular error
$\delta\theta_{\min}$ reached there, for two states of the linear family
and six incident photon numbers. Postselection on $|D\rangle$. For each
state, the coefficient $f(\theta)$ of equation~\eqref{eq:ftheta} and the
postselection probability $\mathcal{N}^2$ are given with the state
itself.}
\label{tab:opt}
\vspace{6pt}
\small
\setlength{\tabcolsep}{5pt}
\begin{tabular}{@{}l l c c c c c c@{}}
\hline
 & & \multicolumn{6}{c}{$N_{\rm total}$}\\
\cline{3-8}
\multicolumn{1}{c}{state} & &
\multicolumn{1}{c}{$10^{6}$} & \multicolumn{1}{c}{$5\times10^{6}$}
 & \multicolumn{1}{c}{$10^{7}$} & \multicolumn{1}{c}{$10^{8}$}
 & \multicolumn{1}{c}{$10^{9}$} & \multicolumn{1}{c}{$10^{10}$}\\
\hline
$\theta=30^\circ$ & $x_{\rm opt}$
 & $0.455$ & $0.348$ & $0.310$ & $0.211$ & $0.144$ & $0.098$\\
$|f(\theta)|=0.116$ & $\eta$
 & $2.6\times10^{-2}$ & $1.5\times10^{-2}$ & $1.2\times10^{-2}$
 & $5.6\times10^{-3}$ & $2.6\times10^{-3}$ & $1.2\times10^{-3}$\\
$\mathcal{N}^2=0.933$ & $\delta\theta_{\min}$ (deg)
 & $0.298$ & $0.174$ & $0.138$ & $0.064$ & $0.030$ & $0.014$\\[4pt]
$\theta=140^\circ$ & $x_{\rm opt}$
 & $0.056$ & $0.043$ & $0.038$ & $0.026$ & $0.018$ & $0.012$\\
$|f(\theta)|=7.236$ & $\eta$
 & $3.9\times10^{-4}$ & $2.3\times10^{-4}$ & $1.8\times10^{-4}$
 & $8.4\times10^{-5}$ & $3.9\times10^{-5}$ & $1.8\times10^{-5}$\\
$\mathcal{N}^2=0.0076$ & $\delta\theta_{\min}$ (deg)
 & $0.219$ & $0.128$ & $0.102$ & $0.047$ & $0.022$ & $0.010$\\
\hline
\end{tabular}
\end{table}

For a small number $N_{\rm total}$, the optimum lies at a stronger
coupling, since the statistical error dominates the systematic one. At
larger $N_{\rm total}$, on the other hand, $\eta$ and
$\delta\theta_{\min}$ both decrease, so a weaker coupling and a better
precision can be reached simultaneously. At $\theta=30^\circ$ and $10^6$
incident photons, $x_{\rm opt}$ equals $0.455$ from
table~\ref{tab:opt}, and the mean arrival time of
equation~\eqref{eq:Texact} is displaced by $0.29\sigma$, close to a third
of the pulse width: the optimal coupling is therefore not weak in the
sense of a small pointer displacement. The expansion of
equation~\eqref{eq:DeltaX} is in $\eta$ and not in $x$, and $\eta$ equals
$2.6$ per cent at that coupling, so the first-order bias and the one
obtained by inverting equation~\eqref{eq:Xq} exactly differ by less than
half a per cent. Equation~\eqref{eq:balance} fixes the residual bias at
$1/\sqrt2$ of the statistical error whatever the value of
$N_{\rm total}$, so this bias does not grow with the photon number. The
exact estimator of section~\ref{sec:exact} removes it, and with it the
constraint from which the optimum follows.

\subsection{The exact estimator}
\label{sec:exact}

The systematic error can nevertheless be avoided, since
equation~\eqref{eq:Xq} can be inverted at fixed $q$ instead of being
expanded. Writing $c=b/a$, the measured quantity becomes
\begin{equation}
X=\frac{1+q\bar c}{1+|c|^2+2q\,\mathrm{Re}(c)},
\label{eq:Xc}
\end{equation}
and, with $D$ denoting that denominator, the inversion becomes the
quadratic equation
\begin{equation}
|X|^2D^2-\left[2\,\mathrm{Re}(X)\left(1-q^2\right)+q^2\right]D
+\left(1-q^2\right)=0,
\label{eq:quadD}
\end{equation}
where $c=\left(\bar XD-1\right)/q$. Of the two roots, the one to be kept
is the one that returns the weak-limit inversion as $q$ tends to unity,
and the state then follows from $c$ through the last two relations of
equation~\eqref{eq:reconstruction}; the estimator built on this inversion
is the exact estimator. An exact reconstruction at finite coupling is
also obtained from two values of the coupling~\cite{Kofman2012} and from
pointer observables that depend on the coupling~\cite{Zhang2016,Zhu2016}.
Here, the mean arrival time and the mean frequency shift of
section~\ref{sec:shifts} carry $\mathrm{Re}(A\bar B)$ and
$\mathrm{Im}(A\bar B)$ at a single value of the coupling, so a single
value is sufficient for the reconstruction and $x$ remains a parameter to
be optimised.

\begin{figure}[htbp]
\centering
\includegraphics[width=0.75\linewidth]{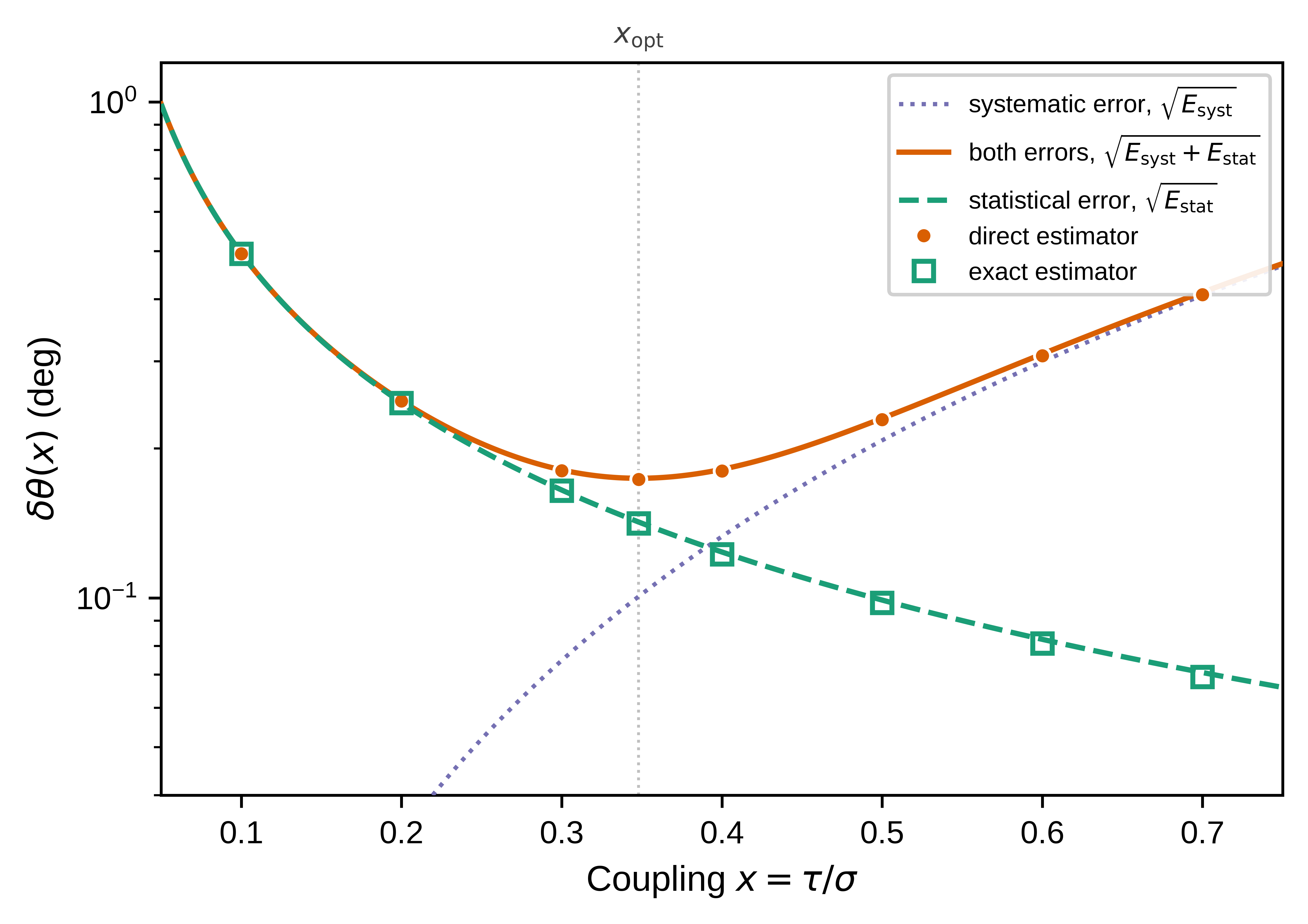}
\caption{\label{fig:estimators}Angular error $\delta\theta$ on the
reconstructed state against the coupling $x$, for the state at
$\theta=30^\circ$ and $N_{\rm total}=5\times10^{6}$ incident photons. The
curves are the closed forms of equation~\eqref{eq:MSE}: the systematic
error alone, the statistical error alone, and the two combined. The
symbols are Monte Carlo results, $2\times10^{5}$ trials per point,
obtained by adding to $X$ the isotropic noise $s$ of
equation~\eqref{eq:s} and applying each estimator to the simulated
records. The direct estimator follows the combined error and passes
through its minimum at $x_{\rm opt}=0.348$, whereas the exact estimator
follows the statistical error alone and decreases over the whole range
shown.}
\end{figure}

The exact inversion removes the systematic error, and the error of the
exact estimator stays within two per cent of the statistical error
$\sqrt{E_{\rm stat}}$ between $x=0.1$ and $x=0.7$, as
figure~\ref{fig:estimators} shows. Since $\sqrt{E_{\rm stat}}$ decreases
as $1/x$, the error of the exact estimator also decreases over that
interval. A Monte Carlo simulation for the state $\theta=30^\circ$ and
$N_{\rm total}=5\times10^{6}$ incident photons gives, at the optimal
coupling $x_{\rm opt}=0.348$, an angular error $\delta\theta$ of
$0.173^\circ$ for the direct estimator against $0.141^\circ$ for the
exact estimator. At $x=0.60$, $\delta\theta$ reaches $0.308^\circ$ for
the direct estimator whereas it falls to $0.081^\circ$ for the exact
estimator. The two estimators coincide below $x=0.2$, where the
systematic error is negligible, and separate before $x_{\rm opt}$ is
reached, so that the direct estimator is already the less precise of the
two at its own optimum. This optimum exists only because the direct
estimator carries a systematic error, the exact estimator having none to
balance. The exact error varies as $N_{\rm total}^{-1/2}$ against
$N_{\rm total}^{-1/3}$ for the optimised direct readout, so the gap
between the two grows with $N_{\rm total}$. If $x$ is known only to ten
per cent, the systematic error remaining at $x=0.5$ is
$\delta\theta=0.043^\circ$, below the statistical error of $0.099^\circ$
at that coupling.

\subsection{The fold of the two-centroid readout}
\label{sec:fold}

\begin{figure}[htbp]
\centering
\includegraphics[width=0.75\linewidth]{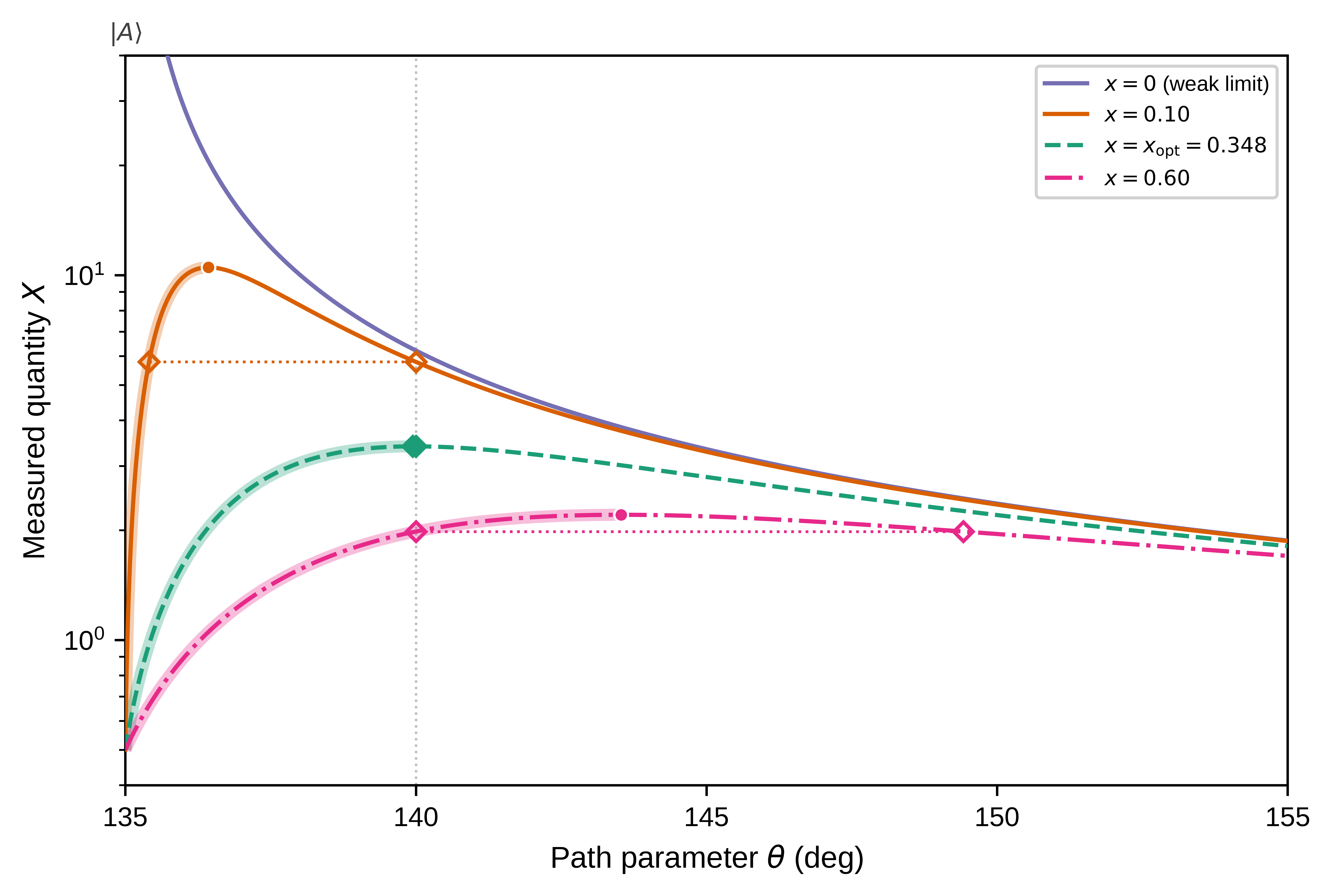}
\caption{\label{fig:fold}Measured quantity $X$ along the family of linear
states near $|A\rangle$, for four values of the coupling. At zero
coupling $X$ decreases monotonically and each measured value corresponds
to a single state. At finite coupling $X$ passes through a maximum,
marked by a dot, and the states lying on either side of that maximum give
the same measured value: the open diamonds mark one such pair, joined by
the horizontal dotted line of the same colour. The shaded part of each
curve runs from $|A\rangle$ to the maximum, and every state it contains
has a twin beyond the maximum. The maximum moves away from $|A\rangle$ as
the coupling increases, and the state at $\theta=140^\circ$ changes side:
outside the fold at $x=0.10$, exactly at the maximum at
$x=x_{\rm opt}=0.348$, and inside the fold at $x=0.60$, where its twin
lies at $149.42^\circ$.}
\end{figure}

The second root of equation~\eqref{eq:quadD}, discarded in
section~\ref{sec:exact} by continuity with the weak limit, is examined
here. Both roots are exact preimages in the neighbourhood of $|A\rangle$,
so that criterion, used in~\cite{Kofman2012}, no longer identifies the
state there, and the map from the state to the pair of centroids is
two-to-one, as figure~\ref{fig:fold} shows; the fold denotes this loss of
injectivity. Along the linear family, the function $X(\theta;q)$
representing the measured quantity is monotonic at $q=1$ and acquires a
turning point when $q<1$, at the postselection amplitude
\begin{equation}
\varepsilon_{\rm fold}=\frac{x}{4},
\label{eq:fold}
\end{equation}
up to a relative correction of $-x^2/32$, below $1.2$ per cent for a
coupling below $0.6$; at $x=0.10$ this point lies $2.9^\circ$ from
$|A\rangle$, that is at $\varepsilon_{\rm fold}=0.025$. This turning
point therefore falls on the floor $\varepsilon_{\rm weak}=x/4$ of
equation~\eqref{eq:eps3}, and the fold is the region between $|A\rangle$
and that point, where every state has a twin beyond it.

The ambiguity concerns the two centroids and not the measurement itself.
At $x=0.10$, the two states $\theta=140^\circ$ and
$\theta=135.409^\circ$ give mean arrival times that differ by
$3.3\times10^{-4}\sigma$, but their profiles have widths of $0.93\sigma$
and $1.60\sigma$, a ratio of $1.7$. A small number of photons therefore
separates them from the pulse shape alone, the two centroids taking the
same value for both states while the profiles differ, as
section~\ref{sec:profile} establishes. For the state at $140^\circ$,
which lies $10^\circ$ from $|A\rangle$ on the great circle $S_3=0$ and
has amplitude $\varepsilon=0.0872$, the second root lies at
$135.409^\circ$ for $x=0.10$, hence short of the prepared state, which is
outside the fold. That second root is then below the floor
$\varepsilon_{\min}$ of section~\ref{sec:cap}, and the ambiguity it
establishes concerns the map and not a measurable state. At
$x=x_{\rm opt}=0.348$, the second root is reached at $139.944^\circ$,
very close to the prepared state, because the edge of the fold,
$\varepsilon_{\rm fold}=0.0867$, nearly equals the amplitude
$\varepsilon=0.0872$ of that state. At $x=0.60$, the second root lies at
$149.42^\circ$, beyond the prepared state, which is then inside the fold.
Above that coupling, the second root falls in the accessible domain and
nothing in the two centroids allows a choice between it and the prepared
state.

Far from $|A\rangle$, the exact estimator can be used and the coupling
increased. At a postselection amplitude $\varepsilon$, the coupling must
stay below $4\varepsilon$. The two estimators are therefore bounded
differently: the direct estimator by the systematic error it accepts,
which fixes $x_{\rm opt}$ of equation~\eqref{eq:xopt}, and the exact
estimator by the amplitude below which the two centroids no longer
determine the state. That amplitude equals $x/4$, so
$\varepsilon_{\rm weak}$ of equation~\eqref{eq:eps3} no longer bounds the
reconstruction but this readout, $\varepsilon_{\rm stat}$ and
$\varepsilon_{\rm det}$ remaining.

%% file: section6_conclusion.tex

\section{Conclusion}
\label{sec:conclusion}

We have developed the theoretical framework for the direct reconstruction of
the state vector of a polarisation state from weak measurements with a
temporal Gaussian pointer, and analysed the errors associated with this
method. The reconstruction formulae established here hold for an arbitrary
postselection state and are exact at zero coupling. They require only the
measurement of two centroids together with a normalisation, so that each
coefficient of the state vector follows from a single reading rather than
from the global inversion required by state tomography.

The states that the method can reconstruct are those lying outside a
spherical exclusion cap of geodesic half-angle $\alpha_{\min}$, centred on
the state $|\varsigma_\perp\rangle$ orthogonal to the postselection state
$|\varsigma\rangle$. For a reconstruction based on the two centroids, the
minimal postselection amplitude $\varepsilon_{\min}$ from which
$\alpha_{\min}$ follows is the largest of three independent thresholds, a
statistical bound on the number of postselected photons, a detector bound and
a weak-regime bound, and it involves no adjustable parameter. Shot noise
imposes none of these bounds, since the amplification of the weak value and
the depletion of the postselected flux cancel exactly, whereas the cap is
closed by the detector bound for a small number of incident photons and by
the weak-regime condition for a large one, where it saturates at $x/4$.

Two states are distinguishable by the reconstruction method only if they
differ by at least the minimal step of equation~\eqref{eq:crit}, which
depends on the coupling, on the number of incident photons and on the
sensitivity of the weak value along the path followed, and involves no
detector parameter. Along the equator the step varies from one state to
another and decreases as the state approaches the excluded cap, where it
stops at the boundary, whereas on the great circle $S_2=0$ the postselection
probability is constant and the step is the same for every state of the path.

For the direct estimator the systematic error grows with the coupling whereas
the statistical error decreases, so that the total error is minimal at an
optimal coupling. This coupling varies as $N_{\mathrm{total}}^{-1/6}$, and
the residual error it leaves varies as $N_{\mathrm{total}}^{-1/3}$. The exact
estimator, by contrast, has no systematic error to balance, so it has no
optimum, and its error follows the statistical error alone, which varies as
$N_{\mathrm{total}}^{-1/2}$; the error of the direct estimator therefore
exceeds it by a factor $N_{\mathrm{total}}^{1/6}$. The exact estimator is
instead bounded by the fold, the region where two states give the same pair
of centroids, which extends up to the same postselection amplitude $x/4$ that
closes the cap.

Using these results, an experimenter designing a direct reconstruction setup
can fix the coupling from a rough estimate of the state, predict the
precision that will be reached, and know which states are out of reach. The
results depend on the delay and the pulse duration only through their ratio
$x$, so that the coupling is set by the pulse duration alone once the delay
is fixed by the set-up, without modifying the optical path. Moving towards
this optimal coupling costs little in accessible states, since going from
$x=0.01$ to $x=0.10$ multiplies the signal-to-noise ratio on the measured
quantity by ten while it multiplies the radius of the excluded cap by 2.5
only.

The postselection state is a second choice left to the experimenter, since
the reconstruction formulae hold for any postselection and the cap is centred
on the state orthogonal to it. Its position is therefore not imposed by the
method, and it can be placed away from the states under study.

A natural extension applies a second delay to the same pointer, so that this
pointer carries two successive weak couplings and gives access to the full
density matrix, as demonstrated with a single read-out degree of freedom
in~\cite{MartinezBecerril2021}. The joint weak value then requires the second
moments of the postselected profiles in addition to their centroids.

%% file: references.bib
@book{vonNeumann1932,
  author    = {von Neumann, John},
  title     = {Mathematische Grundlagen der Quantenmechanik},
  publisher = {Springer},
  address   = {Berlin},
  year      = {1932}
}

@article{AAV1988,
  author  = {Aharonov, Yakir and Albert, David Z. and Vaidman, Lev},
  title   = {How the result of a measurement of a component of the spin
             of a spin-1/2 particle can turn out to be 100},
  journal = {Phys. Rev. Lett.},
  volume  = {60},
  pages   = {1351},
  year    = {1988},
  doi     = {10.1103/PhysRevLett.60.1351}
}

@article{Jozsa2007,
  author  = {Jozsa, Richard},
  title   = {Complex weak values in quantum measurement},
  journal = {Phys. Rev. A},
  volume  = {76},
  pages   = {044103},
  year    = {2007},
  doi     = {10.1103/PhysRevA.76.044103}
}

@article{Lundeen2011,
  author  = {Lundeen, Jeff S. and Sutherland, Brandon and Patel, Aabid
             and Stewart, Corey and Bamber, Charles},
  title   = {Direct measurement of the quantum wavefunction},
  journal = {Nature},
  volume  = {474},
  pages   = {188},
  year    = {2011},
  doi     = {10.1038/nature10120}
}

@article{Salvail2013,
  author  = {Salvail, Jeff Z. and Agnew, Megan and Johnson, Allan S. and
             Bolduc, Eliot and Leach, Jonathan and Boyd, Robert W.},
  title   = {Full characterization of polarization states of light via
             direct measurement},
  journal = {Nature Photon.},
  volume  = {7},
  pages   = {316},
  year    = {2013},
  doi     = {10.1038/nphoton.2013.24}
}

@article{BrunnerSimon2010,
  author  = {Brunner, Nicolas and Simon, Christoph},
  title   = {Measuring small longitudinal phase shifts: weak
             measurements or standard interferometry?},
  journal = {Phys. Rev. Lett.},
  volume  = {105},
  pages   = {010405},
  year    = {2010},
  doi     = {10.1103/PhysRevLett.105.010405}
}

@article{Kedem2012,
  author  = {Kedem, Yaron},
  title   = {Using technical noise to increase the signal-to-noise ratio
             of measurements via imaginary weak values},
  journal = {Phys. Rev. A},
  volume  = {85},
  pages   = {060102},
  year    = {2012},
  doi     = {10.1103/PhysRevA.85.060102}
}

@article{FerrieCombes2014,
  author  = {Ferrie, Christopher and Combes, Joshua},
  title   = {Weak value amplification is suboptimal for estimation and
             detection},
  journal = {Phys. Rev. Lett.},
  volume  = {112},
  pages   = {040406},
  year    = {2014},
  doi     = {10.1103/PhysRevLett.112.040406}
}

@article{KneeGauger2014,
  author  = {Knee, George C. and Gauger, Erik M.},
  title   = {When amplification with weak values fails to suppress
             technical noise},
  journal = {Phys. Rev. X},
  volume  = {4},
  pages   = {011032},
  year    = {2014},
  doi     = {10.1103/PhysRevX.4.011032}
}

@article{Jordan2014,
  author  = {Jordan, Andrew N. and Mart{\'i}nez-Rinc{\'o}n, Juli{\'a}n
             and Howell, John C.},
  title   = {Technical advantages for weak-value amplification: when
             less is more},
  journal = {Phys. Rev. X},
  volume  = {4},
  pages   = {011031},
  year    = {2014},
  doi     = {10.1103/PhysRevX.4.011031}
}

@article{MartinezBecerril2021,
  doi = {10.22331/q-2021-12-06-599},
  title = {Theory and experiment for resource-efficient joint weak-measurement},
  author = {Martinez-Becerril, Aldo C. and Bussi{\`{e}}res, Gabriel and Curic, Davor and Giner, Lambert and A. Abrahao, Raphael and Lundeen, Jeff S.},
  journal = {{Quantum}},
  issn = {2521-327X},
  publisher = {{Verein zur F{\"{o}}rderung des Open Access Publizierens in den Quantenwissenschaften}},
  volume = {5},
  pages = {599},
  month = dec,
  year = {2021}
}

@article{Kofman2012,
  author  = {Kofman, Abraham G. and Ashhab, Sahel and Nori, Franco},
  title   = {Nonperturbative theory of weak pre- and post-selected
             measurements},
  journal = {Phys. Rep.},
  volume  = {520},
  number  = {2},
  pages   = {43--133},
  year    = {2012},
  doi     = {10.1016/j.physrep.2012.07.001}
}

@article{Dressel2014,
  author  = {Dressel, Justin and Malik, Mehul and Miatto, Filippo M.
             and Jordan, Andrew N. and Boyd, Robert W.},
  title   = {Colloquium: Understanding quantum weak values: Basics and
             applications},
  journal = {Rev. Mod. Phys.},
  volume  = {86},
  pages   = {307},
  year    = {2014},
  doi     = {10.1103/RevModPhys.86.307}
}

@article{Hofmann2011,
  author  = {Hofmann, Holger F.},
  title   = {Uncertainty limits for quantum metrology obtained from the
             statistics of weak measurements},
  journal = {Phys. Rev. A},
  volume  = {83},
  pages   = {022106},
  year    = {2011},
  doi     = {10.1103/PhysRevA.83.022106}
}

@article{Kobayashi2014,
  author  = {Kobayashi, Hirokazu and Nonaka, Koji and Shikano, Yutaka},
  title   = {Stereographical visualization of a polarization state using
             weak measurements with an optical-vortex beam},
  journal = {Phys. Rev. A},
  volume  = {89},
  pages   = {053816},
  year    = {2014},
  doi     = {10.1103/PhysRevA.89.053816}
}

@article{Vallone2016,
  author  = {Vallone, Giuseppe and Dequal, Daniele},
  title   = {Strong measurements give a better direct measurement of the
             quantum wave function},
  journal = {Phys. Rev. Lett.},
  volume  = {116},
  pages   = {040502},
  year    = {2016},
  doi     = {10.1103/PhysRevLett.116.040502}
}

@article{Zhang2016,
  author  = {Zhang, Yu-Xiang and Wu, Shengjun and Chen, Zeng-Bing},
  title   = {Coupling-deformed pointer observables and weak values},
  journal = {Phys. Rev. A},
  volume  = {93},
  pages   = {032128},
  year    = {2016},
  doi     = {10.1103/PhysRevA.93.032128}
}

@article{Zhu2016,
  author  = {Zhu, Xuan-Min and Zhang, Yu-Xiang and Wu, Shengjun},
  title   = {Direct state reconstruction with coupling-deformed pointer
             observables},
  journal = {Phys. Rev. A},
  volume  = {93},
  pages   = {062304},
  year    = {2016},
  doi     = {10.1103/PhysRevA.93.062304}
}

@article{Piacentini2018,
  author  = {Piacentini, F. and Avella, A. and Gramegna, M. and Lussana, R.
             and Villa, F. and Tosi, A. and Brida, G. and Degiovanni, I. P.
             and Genovese, M.},
  title   = {Investigating the effects of the interaction intensity in a weak
             measurement},
  journal = {Scientific Reports},
  volume  = {8},
  pages   = {6959},
  year    = {2018},
  doi     = {10.1038/s41598-018-25156-7}
}

@article{Thekkadath2016,
  author  = {Thekkadath, G. S. and Giner, L. and Chalich, Y. and Horton, M. J.
             and Banker, J. and Lundeen, J. S.},
  title   = {Direct measurement of the density matrix of a quantum system},
  journal = {Physical Review Letters},
  volume  = {117},
  pages   = {120401},
  year    = {2016},
  doi     = {10.1103/PhysRevLett.117.120401}
}

@incollection{Altepeter2004,
  author    = {Altepeter, J. B. and James, D. F. V. and Kwiat, P. G.},
  title     = {Qubit quantum state tomography},
  booktitle = {Quantum State Estimation},
  editor    = {Paris, M. and {\v R}eh{\'a}{\v c}ek, J.},
  series    = {Lecture Notes in Physics},
  volume    = {649},
  pages     = {113--145},
  publisher = {Springer},
  address   = {Berlin},
  year      = {2004},
  doi       = {10.1007/978-3-540-44481-7_4}
}

@article{James2001,
  author  = {James, D. F. V. and Kwiat, P. G. and Munro, W. J. and
             White, A. G.},
  title   = {Measurement of qubits},
  journal = {Physical Review A},
  volume  = {64},
  pages   = {052312},
  year    = {2001},
  doi     = {10.1103/PhysRevA.64.052312}
}

@incollection{DAriano2003,
  author    = {D'Ariano, G. M. and Paris, M. G. A. and Sacchi, M. F.},
  title     = {Quantum tomography},
  booktitle = {Advances in Imaging and Electron Physics},
  editor    = {Hawkes, P. W.},
  volume    = {128},
  pages     = {205--308},
  publisher = {Elsevier},
  address   = {Amsterdam},
  year      = {2003},
  isbn      = {9780120147700},
  issn      = {1076-5670},
  doi       = {10.1016/S1076-5670(03)80065-4}
}

@article{LundeenBamber2012,
  author  = {Lundeen, J. S. and Bamber, C.},
  title   = {Procedure for direct measurement of general quantum states using
             weak measurement},
  journal = {Physical Review Letters},
  volume  = {108},
  pages   = {070402},
  year    = {2012},
  doi     = {10.1103/PhysRevLett.108.070402}
}

@article{Hosten2008,
  author  = {Hosten, O. and Kwiat, P.},
  title   = {Observation of the spin {H}all effect of light via weak
             measurements},
  journal = {Science},
  volume  = {319},
  pages   = {787--790},
  year    = {2008},
  doi     = {10.1126/science.1152697}
}

@article{Dixon2009,
  author  = {Dixon, P. B. and Starling, D. J. and Jordan, A. N. and
             Howell, J. C.},
  title   = {Ultrasensitive beam deflection measurement via interferometric
             weak value amplification},
  journal = {Physical Review Letters},
  volume  = {102},
  pages   = {173601},
  year    = {2009},
  doi     = {10.1103/PhysRevLett.102.173601}
}

@article{Pusey2014,
  author  = {Pusey, M. F.},
  title   = {Anomalous weak values are proofs of contextuality},
  journal = {Physical Review Letters},
  volume  = {113},
  pages   = {200401},
  year    = {2014},
  doi     = {10.1103/PhysRevLett.113.200401}
}

@article{Hariri2019,
  author  = {Hariri, A. and Curic, D. and Giner, L. and Lundeen, J. S.},
  title   = {Experimental simultaneous readout of the real and imaginary parts
             of the weak value},
  journal = {Physical Review A},
  volume  = {100},
  pages   = {032119},
  year    = {2019},
  doi     = {10.1103/PhysRevA.100.032119}
}
